\documentclass[review]{elsarticle}
\usepackage{lipsum}
\makeatletter
\def\ps@pprintTitle{%
 \let\@oddhead\@empty
 \let\@evenhead\@empty
 \def\@oddfoot{}%
 \let\@evenfoot\@oddfoot}
\makeatother
\usepackage{hyperref}

\biboptions{authoryear}

\usepackage{amsmath}
\usepackage{amssymb}
\usepackage{hyperref}
\hypersetup{
    colorlinks = true,
    citecolor = {blue}
}
\usepackage{bm}
\usepackage{subcaption}
\usepackage{url}
\begin{document}

\begin{frontmatter}

\title{Bayesian model selection in the $\mathcal{M}$-open setting -- Approximate posterior inference and probability-proportional-to-size subsampling for efficient large-scale leave-one-out cross-validation}

\author{Riko Kelter\corref{mycorrespondingauthor}}
\address{Department of Mathematics, University of Siegen}


\cortext[mycorrespondingauthor]{Correspondence to: Riko Kelter, Walter-Flex-Street 3, 57072 Siegen, Germany}
\ead{riko.kelter@uni-siegen.de}


\begin{abstract}
Comparison of competing statistical models is an essential part of psychological research. From a Bayesian perspective, various approaches to model comparison and selection have been proposed in the literature. However, the applicability of these approaches strongly depends on the assumptions about the model space $\mathcal{M}$, the so-called model view. Furthermore, traditional methods like leave-one-out cross-validation (LOO-CV) estimate the expected log predictive density (ELPD) of a model to investigate how the model generalises out-of-sample, which quickly becomes computationally inefficient when sample size becomes large. Here, we provide a tutorial on approximate Pareto-smoothed importance sampling leave-one-out cross-validation (PSIS-LOO), a computationally efficient method for Bayesian model comparison. First, we discuss several model views and the available Bayesian model comparison methods in each. We then use Bayesian logistic regression as a running example how to apply the method in practice, and show that it outperforms other methods like LOO-CV or information criteria in terms of computational effort while providing similarly accurate ELPD estimates. In a second step, we show how even large-scale models can be compared efficiently by using posterior approximations in combination with probability-proportional-to-size subsampling. We show how to compare competing models based on the ELPD estimates provided, and how to conduct posterior predictive checks to safeguard against overconfidence in one of the models under consideration. We conclude that the method is attractive for mathematical psychologists who aim at comparing several competing statistical models, which are possibly high-dimensional and in the big-data regime.
\end{abstract}


\end{frontmatter}

\section{Introduction}\label{sec:intro}
Bayesian statistics has become increasingly popular among mathematical psychologists \citep{Wagenmakers2010, Ly2016a, Bayarri2016, Rouder2009, Wang2016}. The benefits of the Bayesian approach are the accordance with the likelihood principle \citep{Birnbaum1962, Berger1988a}, conceptual simplicity \citep{Edwards1963}, ease of interpretation and the possibility to obtain inference even for highly complex models due to the advent of modern computational methods \citep{Kelter2020BayesianPosteriorIndices, Kelter2020BMCJasp, Makowski2019, Kruschke2018}. A particular strength of the Bayesian approach is given by the freedom to use optional stopping (that is, stop sampling participants when the data show already overwhelming evidence), independence of the researcher's intentions \citep{Kruschke2018}, and the interpretation of censored data, all of which are consequences of the accordance of the Bayesian paradigm with the likelihood principle \citep{Birnbaum1962, Berger1988a}. Next to hypothesis testing and parameter estimation, Bayesian model comparison is one of the major applications of the Bayesian approach, and the comparison of statistical models is an essential part of psychological research \citep{Edwards1963, Wagenmakers2010, Kruschke2015, Ly2016a, Ly2016, Gronau2019a}. However, the range of techniques proposed to compare competing statistical models in the Bayesian literature is vast, which makes it difficult to select the appropriate method for model comparison and selection based on (i) the assumptions made about the structure of the model space $\mathcal{M}$ and (ii) the computational complexity of the procedure. The latter aspect becomes important in particular when the dimensionality of the model and the sample size becomes large, a situation which is becoming the norm rather the exception in a variety of psychological research \citep{Harlow2016, Kang2019}.

Established methods to compare competing Bayesian models in psychological research include the Bayes factor \citep{jeffreys1961, KassRaftery1995, Robert2008a, Held2018a}, posterior model probabilities, in particular maximum a posteriori (MAP) model selection \citep{Marin2014, Held2014, Piironen2017a} and Bayesian model averaging (BMA) \citep{Hoeting1999, Raftery2003, Barbieri2004, Claeskens2008}. The second category of approaches consists of cross-validation methods. These include $K$-fold cross-validation \cite{Geisser1979}, which is a biased estimate of precise leave-one-out cross-validation (LOO-CV) \citep{Burman1989}, which itself often is referred to as the gold standard for Bayesian model comparison \citep{Vehtari2012, Piironen2017a}. As precise LOO-CV is computationally very costly, analytical approximations (for some specific models) and numerical approximations (in more generality) have been developed \citep{Vehtari2014, Vehtari2015, Vehtari2017}. Additionally, approximations to precise LOO-CV have been developed, which reduce the computational effort both when the sample size is large and the model is potentially high-dimensional, making posterior inference costly \citep{Magnusson2019}.

Another branch of Bayesian model selection approaches is given by various information criteria like the Akaike information criterion (AIC) \citep{Akaike1974}, the Bayesian information criterion (BIC) \citep{Schwarz1978BayesianInformationCriterion}, the deviance information criterion (DIC) \citep{Spiegelhalter2002} and the widely applicable information criterion (WAIC) \citep{Watanabe2009, Watanabe2010}. For an overview and relationships between these see \cite{Stone1977}, \cite{Konishi2008}, \cite{Vrieze2012} and \cite{McElreath2020}, where \cite{Vrieze2012} places emphasis in particular on the comparison of the AIC and BIC for model selection in psychological research.

Approaches to Bayesian model comparison which are efficient in settings with a large number of predictors and a large number of candidate models include the reference predictive method which minimises the Kullback-Leibler divergence between a prespecified reference model and a candidate model \citep{Martini1984, Piironen2017a}, and the projection predictive inference approach of \cite{GoutisRobert1998}, which was extended by \cite{Dupuis2003} and \cite{Piironen2018}.

An often overlooked fact is that the applicability of existing model comparison and selection techniques strongly depends on the assumptions made about the model space $\mathcal{M}$ \citep{Bernardo1994, Vehtari2012, Piironen2017a}. This complicates the selection of an appropriate and computationally efficient model comparison method in practice. Additionally, some of the above-mentioned techniques are approximations or biased estimators of each other, complicating the situation even further. For example, it is well-known that $K$-fold cross-validation is a biased estimator of LOO-CV \citep{Burman1989} which is computationally more efficient. However, if computational efficiency is not needed because the model's complexity and the sample size are only moderate, introducing bias to reduce the variance is superfluous. Also, LOO-CV itself is asymptotically equal to the AIC for ordinary regression models \citep{Stone1977}, but in the massive-data setting with a huge sample size, it may be much more efficient to compute AIC instead of precise LOO-CV. Additionally, AIC and BIC can be derived in the same Bayesian framework just by using different prior probabilities \citep{Burnham2004}, but the optimality of AIC or BIC depends again (for example, in linear regression) on specific assumptions on the model space $\mathcal{M}$ (the ``true`` model must be in the candidate set) as shown by \cite{Yang2005}. As these examples highlight, the fact that the various methods are often intertwined complicates the selection of the appropriate technique in practice. Model comparison is, therefore, an important technique in psychological research -- see also \cite{Wasserman2000}, \cite{Zucchini2000} and \cite{Vrieze2012} -- and the appropriateness of a specific model selection method depends strongly on the assumptions made about the model space $\mathcal{M}$ and the method's computational efficiency.

The plan of the paper is as follows: Section \ref{sec:BayesianModelCompAsExpUtility} briefly introduces Bayesian model comparison from a decision-theoretic perspective, which strives to maximise an expected utility. It is shown that in general, Bayesian model comparison techniques can be unified under this paradigm by selecting the logarithmic score as the utility function, thereby judging the out-of-sample predictive ability of a statistical model under consideration. Section \ref{sec:modelSelectionApproaches} then discusses different perspectives on the model space $\mathcal{M}$, and the possible model selection methods which are available in each model view. Section \ref{sec:psisloo} introduces Pareto-smoothed importance sampling leave-one-out cross-validation (PSIS-LOO-CV) as a general method in the $\mathcal{M}$-open model view, and relates the underlying theory to standard LOO-CV and importance sampling LOO-CV (IS-LOO-CV). Additionally, the theory behind \textit{approximate} LOO-CV as proposed by \cite{Magnusson2019} is detailed, which makes the method applicable even in large-scale models with huge sample sizes and a potentially large number of model parameters. Section \ref{sec:caseStudy} then provides a case study in which logistic regression is used as a running example in the tutorial to showcase how to use approximate PSIS-LOO-CV in practice. It is shown that PSIS-LOO significantly reduces the computational effort needed for Bayesian model comparison, while retaining high accuracy of the ELPD estimates. Section \ref{sec:discussion} concludes by discussing the benefits and limitations of the method.

\section{Bayesian model comparison from a decision-theoretic perspective: Maximising the expected utility and predictive ability}\label{sec:BayesianModelCompAsExpUtility}
Before the different model space views are discussed, this section briefly explains how Bayesian model comparison methods can be united under the decision-theoretic umbrella of the model's expected utility. This perspective enables to quantify the predictive ability of a model by using a specific utility function (the log-score) and to judge how well it generalises beyond the observed sample data.

From a Bayesian perspective, model comparison is interpreted as a decision-theoretic approach which has the goal of maximising the \textit{expected utility} for a utility function $u(M,\cdot)$ when selecting a model $M \subset \mathcal{M}$ from the model space $\mathcal{M}$, which consists of all models under consideration. The expected utility $\bar{u}(M)$ of selecting a model $M$ is given as
\begin{align}
	\bar{u}(M)=\int u(M,\tilde{y})p_t(\tilde{y})d\tilde{y}
\end{align}
where the true but unknown probability distribution generating an observation $\tilde{y}$ is $p_t(\tilde{y})$. A common approach for model comparison is to analyse how well a model predicts new data \citep{Vehtari2012, Piironen2017a}. If a model does not generalise well to new data, it may be argued that the model is inappropriate or at least lacks some important features so that it does not capture the nature of the true underlying data generating process $p_t(\tilde{y})$. A natural candidate for the utility function $u$ which is therefore commonly used is the log score $u(M,\tilde{y})=\log p_M(\tilde{y}|y)$, which leads to the \textit{expected log predictive density} (ELPD):
\begin{align}\label{eq:elpdbar}
	\overline{\text{elpd}}(M)=\int \log p_M(\tilde{y}|y)p_t(\tilde{y})d\tilde{y}
\end{align}
The ELPD for model $M$ can be interpreted as the weighted average of the log predictive density $\log p_M(\tilde{y}_i|y)$ for a new observation for the model $M$, where the weights stem from the true data generating process $p_t(\tilde{y})$. Large values of $\overline{\text{elpd}}(M)$ indicate that the model predicts new observations $\tilde{y}$ well, while small values of $\overline{\text{elpd}}(M)$ show that the model does not generalise well to new data. However, in practice the true probability density $p_t(\tilde{y})$ of course is unknown, which is why leave-one-out cross-validation (LOO-CV) has established itself as a general method for estimating the ELPD for a model $M$ \citep{Vehtari2017}. The idea of LOO-CV is to interpret the already observed data $D=\{x_i,y_i\}_{i=1}^n$, in particular $y:=(y_1,...,y_n)$ as pseudo Monte-Carlo draws from the true data generating process $p_t(\tilde{y})$ and to estimate the integral in equation (\ref{eq:elpdbar}) via the Monte-Carlo average
\begin{flalign}\label{eq:elpd_LOO-CV}
	\overline{\text{elpd}}_{\text{LOO}}(M)&=\frac{1}{n}\sum_{i=1}^n \log p_M(y_i|y_{-i}) \\
	&=\frac{1}{n}\sum_{i=1}^n \log \int p_M(y_i|\theta)p_M(\theta|y_{-i})d\theta\\ 
	&= \frac{1}{n}\text{elpd}_{\text{LOO}}(M)
\end{flalign}
where $p_M(\cdot |y_{-i})$ denotes conditioning on all observations $y_1,...,y_{i-1},y_{i+1},...,y_n$ except for observation $y_i$, and $p_M(\theta|y_{-i})$ is the leave-one-out (LOO) posterior, henceforth only called the LOO posterior. $p_M(\theta|y)$ is the full posterior distribution conditioned on all data $y_1,...,y_n$ and $p_M(y_i|\theta)$ is the model likelihood. For Bayesian model selection, the quantities $\overline{\text{elpd}}_{\text{LOO}}(M)$ and $\text{elpd}_{\text{LOO}}(M)$ are relevant to judge the out-of-sample predictive performance of competing models under consideration.

From a theoretical perspective, LOO-CV has several properties which makes the procedure attractive for Bayesian model comparisons: First, LOO-CV as given in equation (\ref{eq:elpd_LOO-CV}) is a consistent estimate of $\overline{\text{elpd}}(M)$, while other approaches like $K$-fold cross-validation (where the data is divided in $K$ chunks and each chunk is left out successively, instead of leaving out each single observation successively) are biased estimators of the expected log predictive density $\overline{\text{elpd}}(M)$ \citep{Burman1989}. Unbiased estimators for the $\overline{\text{elpd}}(M)$ like the widely applicable information criterion (WAIC) \citep{Watanabe2009, Watanabe2010} are an alternative to LOO-CV, but recent research hints in the direction that LOO-CV is more robust than WAIC in the finite data setting \citep{Vehtari2017}. 

Second, as stressed by \cite{Magnusson2019}, LOO-CV has the appealing property that different utility functions $u(M, \cdot)$ can easily be used in the procedure. Third, hierarchical data structures like leave-one-group-out cross-validation can be taken into account by LOO-CV as shown by \cite{Merkle2019}. Together, these benefits are in particular appealing for psychological research, where different utility functions (for example based on treatment or intervention costs or the expected treatment utility or outcome) could be used, and hierarchical data is often employed in study designs, for example from patients located at different hospital sites, geographic regions or other group-related structures.

However, psychological research also faces massive amounts of data, and model comparison techniques like LOO-CV need to scale in such situations, which are becoming the norm rather than the exception nowadays \citep{Harlow2016, Kang2019}.

The computational costs of Bayesian model comparison via LOO-CV problematically strongly depend on two quantities as shown in table \ref{tab:1}.
\begin{center}
\begin{table}[h!]
\normalsize\sf\centering
Computational costs of LOO-CV\\
\begin{tabular}{lll}
\hline
 & Small $p$ & Large $p$\\
Small $n$ & Moderate & Large\\
Large $n$ & Large & Very large\\
\hline
\end{tabular}
\caption{Computational costs of Bayesian LOO-CV depending on the number of parameters $p$ and the sample size $n$}
\label{tab:1}
\end{table}
\end{center}
If the number of parameters $p$ is small, the computational costs of obtaining the LOO posterior $p_M(\theta|y_{-i})$ -- which in realistic settings most often is obtained via Markov-Chain-Monte-Carlo (MCMC) or Hamiltonian Monte Carlo (HMC) sampling \citep{Wagenmakers2010} -- are only moderate. In small-dimensional posterior distributions HMC algorithms like the No-U-Turn sampler of \cite{Hoffman2014} work efficiently in exploring the posterior. If additionally the sample size $n$ is small, the number of LOO posteriors $p_M(\theta|y_{-i})$ which need to be computed for the calculation of $\overline{\text{elpd}}_{\text{LOO}}(M)$ is small, too. 

When the number of parameters $p$ grows, but sample size $n$ stays small, the computational costs grow by the increased computational needs of the MCMC or HMC algorithms to obtain the LOO posteriors $p_M(\theta|y_{-i})$, but the model still needs to be refit only $n$ times. However, the increased computational effort of obtaining the posterior for large $p$ can be substantial. The reason is given by the fact that very high-dimensional posteriors typically have a complex posterior geometry, which makes it hard for traditional MCMC and HMC algorithms to explore the distribution efficiently \citep{Neal2011, Betancourt2017}.

In the other case when the sample size $n$ grows, but the number of parameters $p$ stays small, MCMC or HMC algorithms can obtain the LOO posteriors $p_M(\theta|y_{-i})$ quickly, but now $n$ of these LOO posteriors needs to be computed, and $n$ is large.

The worst-case scenario happens in the regime of massive high-dimensional data, which amounts to both $p$ and $n$ being large. In this case, fitting a single LOO posterior $p_M(\theta|y_{-i})$ via MCMC or HMC is already costly, and the model needs to be fitted a large number of $n$ times. Unfortunately, this data-rich regime is often faced in psychological research, for example when a substantial number of covariates (e.g. biomarkers) are recorded, and the number of study participants is large, too. Also, the situation in which $n$ is small and $p$ is large is often faced, too. For example, when studying rare disorders the sample size $n$ is typically much smaller than $p$, and the computational costs remain large (see upper right cell of table \ref{tab:1}).

Before we introduce approximate PSIS-LOO-CV as a potential solution to these problems, the following section briefly discusses multiple model views, each of which makes specific assumptions about the model space $\mathcal{M}$ of all models under consideration.


\section{Model views and their corresponding Bayesian approaches to model comparison and selection}\label{sec:modelSelectionApproaches}
This section discusses multiple model views and the corresponding model comparison and selection methods available in each of these views. A comprehensive review of Bayesian model selection perspectives is given by \cite{Bernardo1994} and \cite{Vehtari2012}. Here, we briefly summarise the different perspectives on Bayesian model selection to clarify the assumptions made by the different model views and discuss the implied restrictions and suitability for psychological research.

The three usually separated model perspectives are shown in table \ref{tab:2}. Note that some authors additionally list the $\mathcal{M}$-mixed view \citep{Vehtari2012, Piironen2017a}. We refrain from doing so because the approaches in the $\mathcal{M}$-mixed view conceptually produce biased estimates of the expected log predictive density (or for the expected utility based on some utility function different from the log-likelihood), and strongly resemble the $\mathcal{M}$-closed or $\mathcal{M}$-completed view.

\begin{center}
\begin{table*}[h!]
\normalsize\sf\centering
Bayesian model selection approaches and model perspectives
\begin{tabular}{p{2cm}p{4.5cm}p{4.5cm}}
\hline
Perspective & Interpretation & Model selection approaches\\
\hline
$\mathcal{M}$-open & No candidate model is correct and there exists no reference model & Cross-validation, information criteria \\
$\mathcal{M}$-completed & Existence of a full encompassing reference model which is treated as the best description of future data & Reference predictive method, projection predictive inference\\
$\mathcal{M}$-closed & Collection of models $\{M_l\}_{l=1}^L$ under consideration, where one assumes that one model $M_l$ is correct & Maximum a posteriori (MAP) model selection, Bayesian model averaging (BMA)\\
\hline
\end{tabular}
\caption{Bayesian model selection approaches and model perspectives}
\label{tab:2}
\end{table*}
\end{center}
\subsection{The $\mathcal{M}$-closed view}
The simplest perspective is the $\mathcal{M}$-closed view, in which the true model $M_t$ is assumed to be among the candidate models $\{M_l\}_{l=1}^L$ under consideration. In this case, the model space $\mathcal{M}$ is closed and the natural Bayesian approach of selecting among the models $\{M_l\}_{l=1}^L$ is to calculate the posterior model probabilities
\begin{align}\label{eq:posteriorModelProb}
    p(M|D)\propto p(D|M)p(M)    
\end{align}
after observing the data $D=\{(x_i,y_i)\}_{i=1}^n$. The usual practice is to select the model with maximum a posteriori (MAP) probability in equation (\ref{eq:posteriorModelProb}) from a model selection perspective \citep{Held2014, Marin2014}. Predictions for new data $\tilde{y}$ are usually obtained via \textit{Bayesian model averaging (BMA)} \citep{Hoeting1999}:
\begin{align*}
    p(\tilde{y}|D)=\sum_{l=1}^L p_{M_l}(\tilde{y}|D)p_{M_l}(D)
\end{align*}
which is simply a model-weighted average of the posterior predictive distributions $p_{M_l}(\tilde{y}|D)$ of each model $M_l$ for the new data $\tilde{y}$\footnote{For notational simplicity, we write the posterior distribution $p(\tilde{y}|D,M_l)$ as $p_{M_l}(\tilde{y}|D)$, where the subscript $M$ indicates the model the posterior distribution is conditioned on.}. The average is built from the posterior predictive distributions of all models $\{M_l\}_{l=1}^L$ under consideration, and the weight of each individual posterior predictive contribution $p_{M_l}(\tilde{y}|D)$ is given by the posterior model probability $p_{M_l}(D)$. For details about BMA see also \cite{Hoeting1999} and \cite{Raftery2003}. While conceptually straightforward, the $\mathcal{M}$-closed assumption is often strongly unrealistic in practice, because (1) it remains entirely unclear if the true model $M_t$ is indeed among the set of candidate models and (2) in a large number of settings the number of models to be considered is infinite (for example in a standard normal model where data is assumed to follow a $\mathcal{N}(\mu,\sigma^2)$ distribution with known $\sigma^2 >0$ and unknown mean $\mu$, so that the number of models to be considered is not countable anymore), which makes the calculation of Bayesian model averages difficult.

\subsection{The $\mathcal{M}$-completed view}
The $\mathcal{M}$-completed view weakens the unrealistic assumption of the $\mathcal{M}$-closed view by assuming only the existence of a \textit{reference model} $M_{\text{ref}}$. This reference model $M_{\text{ref}}$ is defined as a full encompassing model over the other candidate models and is believed to represent the underlying reality best among all models under consideration. However, it is not believed to be the true data generating model $M_t$, and also the assumption that the model space $\mathcal{M}$ of models under considerations contains the true data generating model $M_t$ is not made in the $\mathcal{M}$-completed view. There are two distinct approaches to the $\mathcal{M}$-completed view: (1) The reference predictive method \citep{Martini1984, Piironen2017a} and (2) projection predictive inference \citep{GoutisRobert1998, Dupuis2003, Piironen2018}.

\subsubsection*{Reference predictive method}
The reference predictive method proceeds by (1) constructing the reference model $M_{\text{ref}}$, and (2) estimating the utilities of the candidate models $M$ under consideration. Notice that in most settings the reference model may be best in terms of predictive ability, but simultaneously difficult to interpret because it has a much larger number of parameters. The goal of the reference predictive method therefore is to \textit{treat} the reference model $M_{\text{ref}}$ as the true underlying data generating distribution $p_t(\tilde{y})$ in equation (\ref{eq:elpdbar}). While it is not believed to be the true data generating model, interpreting it as if it were the true model then enables to use it as a benchmark to compare the other candidate models with it in terms of predictive ability. These other candidate models are in general easier to interpret than the reference model $M_{\text{ref}}$, and if a candidate model $M$ can be found which yields a similar predictive ability as the reference model $M_{\text{ref}}$, this simplifies model interpretation.

The expected utility using the reference model $M_{\text{ref}}$ is obtained by replacing the true distribution $p_t(\tilde{y})$ in equation (\ref{eq:elpdbar}) with the reference model's predictive distribution: $p_t(\tilde{y})=p_{M_{\text{ref}}}(\tilde{y}|D)$. Computing the average over the covariate inputs $\{x_i\}_{i=1}^n$ of the training data $D$ yields the reference utility $\bar{u}_{\text{ref}}(M)$:
\begin{align}\label{eq:expectedReferenceUtility}
    \bar{u}_{\text{ref}}(M)=\frac{1}{n}\sum_{i=1}^n \int \log p_{M_{\text{ref}}}(\tilde{y}|x_i,D) p_{M}(\tilde{y}|x_i,D)d\tilde{y}     
\end{align}
The maximisation of this expected utility is equivalent to minimisation of the predictive Kullback-Leibler divergence between the candidate model $M$ and the reference model $M_{\text{ref}}$ at the data:
\begin{align}
    \delta(M_{\text{ref}}||M)=\frac{1}{n}\sum_{i=1}^n \text{KL}(p_{M_{\text{ref}}}(\tilde{y}|x_i,D) || p_{M}(\tilde{y}|x_i,D)) 
\end{align}
However, while the model selection can be based on selecting the model $M_s$ which minimises $\delta(M_{\text{ref}}||M)$, that is $M_s := \text{arg} \min\limits_{M} \delta(M_{\text{ref}}||M)$, there are multiple challenges to this approach. First, the computation of the expected utility in equation (\ref{eq:expectedReferenceUtility}) can become difficult, in particular when $\tilde{y}$ is high-dimensional. Second, the construction of the reference model $M_{\text{ref}}$ is problematic in practice, as in most situations it simply is \textit{``not obvious how it should be done.''} \cite[p.~715]{Piironen2017a}. While \cite{Martini1984} used the Bayesian model average as a default reference model $M_{\text{ref}}$, computing the BMA may be difficult in practice, too. Even worse, the BMA assumes the $\mathcal{M}$-closed view, which is an unrealistic assumption in most situations.

\subsubsection*{Projection predictive inference}
The second option in the $\mathcal{M}$-completed view is the projection predictive inference approach. The approach goes back to \cite{GoutisRobert1998} and \cite{Dupuis2003}, and the general idea is that the information contained in the reference model $M_{\text{ref}}$ is projected onto the candidate models $M$ to cause the predictive distribution of the reference and candidate model to be as similar as possible. The proposal of \cite{GoutisRobert1998} was to project the parameter $\theta_{*}$ in the parameter space $\Theta_{M_{\text{ref}}}$ of the reference model $M_{\text{ref}}$ onto a parameter $\theta_{M}$ in the parameter space $\Theta_{M}$ of the candidate model $M$. The projection is then defined as the minimisation of the Kullback-Leibler divergence between the predictive model densities over the parameters $\theta$:
\begin{align}\label{eq:projectedParameter}
    \theta_{M}=\arg \min\limits_{\theta} \frac{1}{n}\sum_{i=1}^n \text{KL}(p_{M_{\text{ref}}}(\tilde{y}|x_i,\theta_{*}) || p_{M}(\tilde{y}|x_i,\theta))
\end{align}
After identifying the parameter $\theta_{M}$ which minimises the Kullback-Leibler divergence between the predictive model densities, the KL-divergence is then used a second time to measure the discrepancy between the reference model $M_{\text{ref}}$ and the candidate model $M$. Now, the parameters in the posterior predictive distribution $p_{M}(\tilde{y}|x_i,\theta)$ of the candidate model $M$ are set to the (optimal) projected parameter $\theta_{M}$, and the KL-divergence is calculated:
\begin{align}\label{eq:projectionPredictiveKL}
    \delta(M_{\text{ref}}||M)=\frac{1}{n}\sum_{i=1}^n \mathbb{E}\left [ \text{KL}(p_{M_{\text{ref}}} (\tilde{y}|x_i,\theta_{*}) || p_{M}(\tilde{y}|x_i,\theta_{M}))\right ]
\end{align}
\cite{Dupuis2003} sampled the posterior of the reference model $M_{\text{ref}}$ and calculated (\ref{eq:projectedParameter}) for each posterior draw $\theta_{*}^s$, $s=1,...,S$, yielding corresponding projected parameters $\theta_M^s$, $s=1,...,S$. Equation (\ref{eq:projectionPredictiveKL}) was then estimated via simple Monte-Carlo integration. Finally, the Monte Carlo estimator of $\delta(M_{\text{ref}}||M)$ in (\ref{eq:projectionPredictiveKL}) can then be used for model selection.

However, the approach has several drawbacks: First, the term (\ref{eq:projectionPredictiveKL}) needs to be obtained numerically in most settings. Second, the optimisation in (\ref{eq:projectedParameter}) also needs to be carried out by numerical methods. Third, \cite{Dupuis2003} used a measure called \textit{relative explanatory power} for model selection, which is not without problems as (1) it uses an arbitrary threshold for deciding between models and (2) interpretation of differences in explanatory power between models is complicated. Details are omitted here due to space limitations but more information can be found in \cite{Nott2009} \cite{Piironen2017a}. Piironen et al. \cite{Piironen2018} recently improved the approach by mapping posterior draws of the reference model onto clusters of parameter values in the candidate model space $\Theta_M$, and highlighted how to use this method for variable selection in the generalised linear model \citep{McCullagh1989}. However, while promising, the construction of the reference model $M_\text{ref}$ remains largely unanswered as simply a reference model is selected in their approach which is best in predictive ability. While this uncouples model selection from variable selection and parameter estimation, in practice a reference model (if one exists) would be based on domain-specific knowledge or prior information, and rarely be justified on purely quantitative arguments.

\subsection{The $\mathcal{M}$-open view}
The most general view is the $\mathcal{M}$-open view, which assumes that none of the models under consideration is true. Also, one assumes that there is no reference model available, which is a full encompassing model for all other models and is assumed to model the underlying reality best. This is the most realistic assumption in most psychological research, and the predictive performance of a model in the $\mathcal{M}$-open view is estimated via cross-validation (CV) or information criteria.

\subsubsection*{Cross-validation}
As already stressed in section \ref{sec:BayesianModelCompAsExpUtility}, the true data generating distribution $p(\tilde{y})$ is unknown, so that one needs to estimate it to obtain the expected log predictive density in equation (\ref{eq:elpdbar}). While it would be possible to use the already observed data $y$ as a proxy for the true data generating distribution $p(\tilde{y})$, using the same data which was used to fit the model itself also to estimate the expected log predictive density will lead to an overly optimistic estimate of the generalisation performance of the model. Therefore, the already observed data $D=\{(x_i,y_i)\}_{i=1}^n$ is divided into $K$ subsets $D_1,...,D_K$, and each of these sets is used for validation of the model which is trained on the remaining subsets. This is the idea of Bayesian $K$-fold cross-validation, which is computed as
\begin{align}
    \overline{\text{elpd}_{\text{K-fold}}}(M)=\frac{1}{n}\sum_{i=1}^n \log p_M(y_i|x_i,D_{-{y_i}})    
\end{align}
Here, $D_{-{y_i}}$ is the training set which includes all partitions $D_i$, $i=1,...,K$, except for the partition with $y_i \subseteq D_i$ in which the observation $y_i$ is located. When setting $K=n$, $K$-fold cross-validation equals leave-one-out cross-validation as detailed in equation (\ref{eq:elpd_LOO-CV}). While $K$-fold CV is a biased estimator of precise LOO-CV \citep{Burman1989}, it is computationally much more efficient than precise LOO-CV when $K<<n$. In these cases, refitting the model needs only to be done $K$ instead of $n$ times. Also, due to the well-known bias-variance-tradeoff \citep{Held2014}, introducing bias may decrease the mean squared error (MSE) and out-of-sample predictive accuracy. This is why $K$-fold CV, in general, is quite robust and precise LOO-CV -- which is an unbiased estimate of the expected log predictive density, see \cite{Watanabe2010} -- may tend to overfitting if the number of considered models becomes large \citep{Vehtari2017, Piironen2018}.

\subsubsection*{Information criteria}
Information criteria are the second option when adopting the $\mathcal{M}$-open perspective. The most popular Bayesian information criterion is the widely applicable information criterion (WAIC) which was introduced by \cite{Watanabe2009, Watanabe2010}. WAIC is given as
\begin{align}\label{eq:waic}
    \text{WAIC}(M)=\frac{1}{n}\sum_{i=1}^n \log p_M(y_i|x_i,D)-\frac{V}{n}    
\end{align}
where the first term is the expected log predictive density and the latter term is the functional variance $V$ scaled by the sample size $n$. $V$ itself can be calculated as
\begin{flalign}
    V=\sum_{i=1}^n &( \mathbb{E} \left [ (\log p_M(y_i|x_i,\theta))^2 \right ] 
    -\mathbb{E} \left [ (\log p_M(y_i|x_i,\theta)) \right ]^2     )
\end{flalign}
In the above, the expectation $\mathbb{E}$ is taken over the posterior $p_M(\theta|D)$. Watanabe \cite{Watanabe2010} showed that the WAIC is asymptotically identical to Bayesian LOO-CV and therefore an unbiased estimator of the expected log predictive density $\overline{\text{elpd}}(M)$ in equation (\ref{eq:elpdbar}). The deviance information criterion (DIC) proposed by \cite{Spiegelhalter2002} is another popular information criterion:
\begin{align}
    \text{DIC}(M)=\frac{1}{n}\sum_{i=1}^n \log p_M(y_i|x_i,D,\bar{\theta})-\frac{p_e}{n}
\end{align}
Here, $p_e$ is an estimate of the effective number of parameters and $\bar{\theta}=\mathbb{E}[\theta|M,D]$ is the posterior mean of the parameters. In particular, this latter assumption can hardly be justified, as it is unclear why the model fit should be computed only for the parameters fixed at the posterior mean. Also, this makes DIC a biased estimate of the expected log predictive density $\overline{\text{elpd}}(M)$. Together, these aspects make DIC a less attractive option compared to WAIC (although the first criticism plays a much more important role; for example, $K$-fold CV is also a biased estimator but enjoys very desirable robustness \citep{Piironen2017a}). For a discussion why to prefer WAIC instead of the AIC \citep{Akaike1974} or BIC \citep{Schwarz1978BayesianInformationCriterion}, see \cite{McElreath2020}.

\subsection{Model views in psychological research}
Based on the three model views given in table \ref{tab:2}, the question is which model view seems most reasonable in psychological research. The most reductionist model view is the $\mathcal{M}$-closed view, which assumes that one of the models $M_l$ in the set of considered models $\{M_l\}_{l=1}^L$ is true. While the $\mathcal{M}$-completed view seems more restrictive, in many situations it remains unclear how to select the reference model $M_{\text{ref}}$. This fact holds in particular when the number of predictors $p$ becomes large, and there is no causal model relating the predictors to each other. The model view posing the fewest restrictions is the $\mathcal{M}$-open view. In a variety of psychological research, it is entirely unrealistic that one of the models considered precisely reflects the true data generating mechanism and most statistical models can at best be interpreted only as approximations to reality in these cases. Additionally, in most of these settings, there is no full encompassing reference model as it is not clear that the reference model needs to be \textit{encompassing}. For example, when only two predictors are studied on a response variable, the full encompassing model would be the model which includes both predictors (and the interaction effect between them). However, there is no reason to assume in general that the encompassing model using both predictors (and the interaction effect between them) is preferable to the models including only one of both predictors in terms of predictive accuracy. One of the predictors could be pure noise without any effect on the response variable. Clearly, the reference model in such a case would not be the full encompassing model, but the model which includes only the relevant predictor(s).

In summary, the $\mathcal{M}$-open view assumes the most realistic structure about the model space $\mathcal{M}$ and the models under consideration. In situations where a meaningful reference model $M_{\text{ref}}$ can be constructed, the $\mathcal{M}$-completed approach may be suitable. However, in these cases, the application of the reference predictive method or the projection predictive method remains complicated. There are multiple proposals on how to project from the reference model parameter space onto the candidate model parameter space \cite{Piironen2018}, and both numerical optimisation and integration is needed. Also, there are multiple proposals for the measure which is used how to decide between the projected models \citep{Piironen2017a}. In contrast, when adopting the $\mathcal{M}$-open view, cross-validation and information criteria for model comparison and selection are more straightforward to compute. Additionally, cross-validation enables to use customised utility functions $u(M,\cdot)$, which is particularly useful in psychological research. Also, as shown by \cite{Merkle2019}, generalisations of LOO-CV such as leave-one-cluster-out or leave-one-group-out are suitable for studies conducted at several hospital sites or institutions. While there has no equivalent theory been developed for information criteria, this is a substantial benefit of cross-validation and the $\mathcal{M}$-open view.

\section{Pareto-smoothed importance samling leave-one-out cross-validation}\label{sec:psisloo}
This section now introduces approximate Pareto-smoothed importance sampling LOO-CV (approximate PSIS-LOO-CV) as an improvement over standard LOO-CV, importance sampling LOO-CV (IS-LOO-CV), and Pareto-smoothed importance sampling LOO-CV (PSIS-LOO-CV). Standard LOO-CV was detailed in section \ref{sec:BayesianModelCompAsExpUtility}, and here, IS-LOO-CV is detailed first. Then, PSIS-LOO-CV is discussed which improves upon IS-LOO-CV, and subsequently, approximations to PSIS-LOO-CV are introduced. These enable the application of PSIS-LOO-CV even in large-scale models.

\subsection{Importance sampling and Pareto-smoothed importance sampling LOO-CV}
As discussed in section \ref{sec:BayesianModelCompAsExpUtility}, naively implementing LOO-CV would imply that inference needs to be repeated $n$ times for each model under consideration. This quickly becomes costly, and \cite{Gelfand1996} originally proposed to use importance sampling to solve this problem as follows: He estimated $p_M(y_i|y_{-i})$ in equation (\ref{eq:elpd_LOO-CV}) using the importance sampling approximation
\begin{align}\label{eq:importanceSamplingApproximation}
    \log \hat{p}(y_i|y_{-i})=\log \left ( \frac{\frac{1}{S}\sum_{s=1}^S p_M(y_i|\theta_s)w(\theta_s)}{\frac{1}{S}\sum_{s=1}^S w(\theta_s)} \right )
\end{align}
where $w(\theta_s)$ are the importance weights (or ratios) which are given as
\begin{align}\label{eq:importanceWeights}
    w(\theta_s)=\frac{p_M(\theta_s|y_{-i})}{p_M(\theta_s|y)}\propto \frac{1}{p_M(y_i|\theta_s)}    
\end{align}
and $\theta_s$, $s=1,...,S$ are draws from the full posterior distribution $p_M(\theta|y)$. Using importance sampling as proposed by \cite{Gelfand1996}, only the full posterior distribution needs to be computed to conduct LOO-CV. In contrast, using standard LOO-CV, obtaining the $n$ LOO posteriors in general is much more costly. Problematically, the variance of the importance ratios $w(\theta_s)$ can become infinite, invalidating the procedure.

 Therefore, \cite{Vehtari2015} introduced Pareto-smoothed importance sampling (PSIS) where a generalised Pareto distribution is fit to the largest weights $w(\theta_s)$ and these importance sampling ratios are replaced with order statistics from the fitted generalised Pareto distribution. This method reduces the variance of the importance ratios by introducing a small bias. Also, the shape parameter $\hat{k}$ from the generalised Pareto distribution can be used as an indicator how reliable the estimate of $\hat{p}(y_i|y_{-i})$ is, and empirical results show that data-points $y_i$ yielding $\hat{k}>.7$ signal that the estimate of $\log \hat{p}(y_i|y_{-i})$ \textit{can} be unreliable \citep{Vehtari2015}. In such cases, standard cross-validation should be used for the problematic observation $y_i$, so that $\log p_M(y_i|y_{-i})$ is not estimated via $\log \hat{p}(y_i|y_{-i})$, but is calculated (numerically) as $\int p_M(y_i|\theta)p_M(\theta|y_{-i})d\theta$. Of course, when the number of observations with $\hat{k}$-values larger than $.7$ grows, PSIS-LOO-CV becomes quickly inefficient. In the worst case, all observations yield values of $\hat{k}>.7$, and in this situation, PSIS-LOO-CV becomes LOO-CV, as all LOO posteriors need to be obtained in the standard way. In cases where PSIS-LOO-CV faces a lot of problematic $\hat{k}$-values, using $K$-fold CV is recommended instead. An appealing feature of the procedure is that it is able to signal problems when it does not work properly through the Pareto $\hat{k}$ diagnostic values.

\subsection{Scalable approximate PSIS-LOO-CV}
\cite{Magnusson2019} recently proposed to improve on naive Bayesian LOO-CV by targeting the two problems discussed in section \ref{sec:BayesianModelCompAsExpUtility}: (1) Obtaining the posterior when $p$ is large is costly. (2) In the domain where the sample size $n$ is large, refitting each model under consideration $n$ times is costly.

\subsubsection{Using posterior approximations instead of precise MCMC inference} 
To solve (1), \cite{Magnusson2019} proposed to use a posterior approximation $q_M(\theta|y)$ for a model $M$ as the proposal distribution in an importance sampling scheme, where the target distribution is the approximate LOO posterior $p_M(\theta|y_{-i})$ of interest. To ensure that the importance sampling scheme still works, it needs to be corrected for using such a posterior approximation instead of the precise posterior. They adapted the importance ratios $w(\theta_s)$ in equation (\ref{eq:importanceWeights}) to
\begin{align}
    w(\theta_s)&=\frac{p_M(\theta_s|y_{-i})}{q_M(\theta_s|y)}
    =\underbrace{\frac{p_M(\theta_s|y_{-i})}{p_M(\theta_s|y)}}_{=:A}\underbrace{\frac{p_M(\theta_s|y)}{q_M(\theta_s|y)}}_{=:B}\\
    &\propto \frac{1}{p_M(y_i|\theta_s)}\frac{p_M(\theta_s|y)}{q_M(\theta_s|y)}\label{eq:adaptedImportanceSampling}
\end{align}
The idea behind this is intuitive: The importance weights are the product the two terms A and B. The quantity A is the correction from the approximate full posterior $p_M(\theta_s|y)$ to the approximate LOO posterior $p_M(\theta_s|y_{-i})$, while the quantity B can be interpreted as the correction from the posterior approximation $q_M(\theta_s|y)$ to the approximate full posterior $p_M(\theta_s|y)$. To reduce the computational effort, first, only an \textit{approximate} full posterior distribution $p_M(\theta|y)$ is computed, for example via Laplace approximation or variational inference. 

In Laplace approximation, the posterior is approximated where $q_M(\theta|y)$ is a multivariate normal distribution and the posterior mode is used as its mean and the inverse Hessian is used as the covariance matrix \citep{Azevedo-Filho1994}.

In variational inference, the Kullback-Leibler (KL) divergence is minimised between an approximate family of densities $\mathcal{F}$ and the posterior distribution $p(\theta|y)$, yielding an approximation which is closest to the true posterior where the distance is measured by the KL divergence \citep{Jordan1999, Kucukelbir2015, Blei2017}. Often, the family of distributions considered is also multivariate normal distributions with either diagonal covariance structure (mean-field) or a full covariance structure (full-rank) \citep{Kucukelbir2015}.

No matter which posterior approximation technique is used, the \textit{approximate} full posterior distribution $p_M(\theta|y)$ is evaluated at $\theta_s$, for $s=1,....,S$ different draws, yielding $p_M(\theta_s|y)$. For Laplace approximation or variational inference, $q_M(\theta_s|y)$ is also known and can be evaluated at the posterior draws $\theta_s$, and the same is true for the likelihood $p_M(y_i|\theta_s)$.

In summary, the calculation of $w(\theta_s)$ via equation (\ref{eq:adaptedImportanceSampling}) is therefore possible based only on a posterior \textit{approximation} instead of full MCMC inference.

Now, plugging the importance weights $w(\theta_s)$ into equation (\ref{eq:importanceSamplingApproximation}) then yields an importance sampling approximation $\log \hat{p}(y_i|y_{-i})$ of $p_M(y_i|y_{-i})$ in equation (\ref{eq:elpd_LOO-CV}), where equation (\ref{eq:elpd_LOO-CV}) itself the standard pseudo Monte-Carlo estimate $\overline{\text{elpd}}_{\text{LOO}}(M)$ for the expected log predictive density $\overline{\text{elpd}}(M)$, which quantifies how well the model generalises beyond the sample.

Like in regular importance sampling, the importance ratios $w(\theta_s)$ here can become unstable due to a long right tail \citep{Gelfand1996}. Therefore, \cite{Magnusson2019} proposed to use PSIS to stabilise the importance ratios and to evaluate the reliability of the posterior approximation using $\hat{k}$ as a diagnostic. This is possible because in cases where the posterior approximation is bad (and $q_M(\theta_s|y)$ in turn becomes small), the ratios $w(\theta_s)$ will become inflated quickly, leading to large $\hat{k}$ values.

While the original proposal of \cite{Gelfand1996} required to compute the full posterior to obtain the importance sampling approximation $\log \hat{p}(y_i|y_{-i})$ for $p_M(y_i|y_{-i})$ in equation (\ref{eq:elpd_LOO-CV}), now only a posterior \textit{approximation} is needed. This decreases the computational burden in the setting when the number of predictors $p$ is large, as Markov-Chain-Monte-Carlo (MCMC) and Hamiltonian Monte Carlo (HMC) algorithms then often encounter problems in efficiently exploring the posterior or the exploration takes quite long \citep{Betancourt2017}. 

\subsubsection{Using probability-proportional-to-size subsampling}
When $n$ is large, the estimation of $p_M(y_i|y_{-i})$ by $\log \hat{p}(y_i|y_{-i})$ with the adapted importance sampling scheme for each term in $\overline{\text{elpd}}_{\text{LOO}}(M)$ can still be costly, as $n$ LOO posteriors need to be derived via importance sampling. \cite{Magnusson2019} therefore additionally proposed to make use of \textit{probability-proportional-to-size subsampling}, which reduces the cost of computing the expected log predictive density $\overline{\text{elpd}}_{\text{LOO}}(M)$ additionally . Instead of estimating $p_M(y_i|y_{-i})$ by $\log \hat{p}(y_i|y_{-i})$ with the adapted importance sampling scheme for each $y_i$, one samples $m<n$ observations proportional to $\tilde{\pi}_i \propto \pi_i =-\log p_M(y_i|y)=-\log \int p_M(y_i|\theta)p_M(\theta|y)d\theta$. Often, the full posterior log predictive density $\log p_M(y_i|y)$ can be computed easily when evaluating models. Even if the computation is costly or complicated, one can make use of the Laplace (or variational inference) approximation in regular models: For large sample size $n$, $\log p_M(y_i|y)\approx \log p_M(y_i|\hat{\theta})$, where $\hat{\theta}$ is the posterior mean obtained via Laplace approximation (or variational inference). Finally, the expected log predictive density then can be estimated as
\begin{align}\label{eq:probpropEstLOOCV}
    \hat{\overline{\text{elpd}}}_{\text{LOO}}(M)=\frac{1}{n}\frac{1}{m}\sum_{i=1}^m \frac{1}{\tilde{\pi}_i}\log \hat{p}(y_i|y_{-i})    
\end{align}
where each $\log \hat{p}(y_i|y_{-i})$ is computed via the importance sampling approximation in equation (\ref{eq:importanceSamplingApproximation}), and the ratios are calculated from the adapted importance sampling scheme shown in equation (\ref{eq:adaptedImportanceSampling}). Each $\log \hat{p}(y_i|y_{-i})$ is an estimate for the corresponding $p_M(y_i|y_{-i})$ in equation (\ref{eq:elpd_LOO-CV}). The Monte-Carlo estimator $\overline{\text{elpd}}_{\text{LOO}}(M)$ in equation (\ref{eq:elpd_LOO-CV}) itself estimates the expected log predictive density $\overline{\text{elpd}}(M)$ given in equation (\ref{eq:elpdbar}). In equation (\ref{eq:probpropEstLOOCV}), $m$ is the subsample size used and $\tilde{\pi}_i$ is the probability of subsampling observation $y_i$. In summary, the estimator in equation (\ref{eq:probpropEstLOOCV}) benefits from the adapted importance sampling, so that only a posterior approximation is required, and also needs only $m<n$ summands for estimating $\overline{\text{elpd}}(M)$. If $n$ is large, this can reduce the computational burden to perform PSIS-LOO-CV substantially.

\cite{Magnusson2019} also derived a closed form expression for the variance of the estimator $\hat{\overline{\text{elpd}}}_{\text{LOO}}(M)$, so that using the already obtained quantities, the variance $\mathbb{V}(\hat{\overline{\text{elpd}}}_{\text{LOO}}(M))$ of the estimator $\hat{\overline{\text{elpd}}}_{\text{LOO}}(M)$ can be calculated as follows:
\begin{align}\label{eq:subsamplingError}
    &\mathbb{V}(\hat{\overline{\text{elpd}}}_{\text{LOO}}(M)=\frac{1}{n^2 m(1-m)}\sum_{i=1}^m \left ( \frac{\log \hat{p}(y_i|y_{-i})}{\tilde{\pi}_i}-n \cdot \hat{\overline{\text{elpd}}}_{\text{LOO}}(M)\right )^2
\end{align}
Henceforth, this variance will be called the \textit{subsampling error}. \cite[Proposition 1]{Magnusson2019} showed under quite general assumptions that the subsampling error $\mathbb{V}(\hat{\overline{\text{elpd}}}_{\text{LOO}}(M))$ converges in probability to zero for fixed subsampling size $m$ and posterior draw size $S$ when $n\rightarrow \infty$, for \textit{any} consistent posterior approximation technique.\footnote{Notice that in the limit, a subsample of size $m=1$ in combination with a single posterior draw $S=1$ suffices when $n\rightarrow \infty$ to let the subsampling error decrease to zero. However, in practice increasing $m$ and $S$ will be much easier than increasing sample size $n$.}

However, to compare competing models the variance $\mathbb{V}[\overline{\text{elpd}}_{\text{LOO}}(M)]$ of $\overline{\text{elpd}}_{\text{LOO}}(M)$ as given in equation (\ref{eq:elpd_LOO-CV}) is required, henceforth denoted as $\sigma_{\text{LOO}}^2(M)$ (that is, $\sigma_{\text{LOO}}^2(M):=\mathbb{V}[\overline{\text{elpd}}_{\text{LOO}}(M)]$). Based on the derivations in \citep{Magnusson2019}, one can use the same observations as sampled before to estimate this variance as follows:
\begin{align}\label{eq:elpdErrorModelComparison}
    &\hat{\sigma}_{\text{LOO}}^2(M)=\frac{1}{nm}\sum_{i=1}^m \frac{[\log \hat{p}(y_i|y_{-i})]^2}{\tilde{\pi}_i} \\
    &+\frac{1}{n^2 m(m-1)}\sum_{i=1}^m \left ( \frac{ \log \hat{p}(y_i|y_{-i})}{\tilde{\pi}_i} -\frac{1}{m} \sum_{i=1}^m \frac{ \log \hat{p}(y_i|y_{-i})}{\tilde{\pi}_i}\right )^2 \nonumber\\
    &- \left ( \frac{1}{nm}\sum_{i=1}^m \frac{ \log \hat{p}(y_i|y_{-i})}{\tilde{\pi}_i}\right )^2 \nonumber
\end{align}
For a proof of the unbiasedness of the estimator $\hat{\sigma}_{\text{LOO}}^2(M)$ for $\sigma_{\text{LOO}}^2(M)$ see the supplementary material in \cite{Magnusson2019}.


\section{Case study}\label{sec:caseStudy}
This section now shows how to apply different LOO-CV methods in practice, in particular approximate PSIS-LOO-CV. An R script including all code to reproduce all results and figures can be found at the Open Science Foundation under \url{https://osf.io/qus6j/}.\footnote{However, we do not encourage the reader to execute code parallel to reading, as calculation of several results may vary from seconds to multiple hours.}

\subsection*{Model}
As a running example, we use the standard logistic regression model in which the likelihood of a single observation is given as
\begin{align}
        f(y_i|p)={n \choose y_i}p^{y_i} (1-p)^{n-y_i}
\end{align}
where $p = l^{-1}(\eta)$ is the probability of a success, $\eta = \alpha + \bm{x}^T \bm{\beta}$ is the linear predictor and $n$ is the number of trials conducted. For the complete sample, the likelihood becomes the product of the single likelihood contributions of each observation. The standard choice to map the predictor $\eta$ which typically is $\in \mathbb{R}$ onto the probability scale $[0,1]$ is to use the logit function $l(p)=\ln(\frac{p}{1-p})$, because then $p=l^{-1}(\eta)=\frac{e^\eta}{1+e^\eta}$. The likelihood of a single observation thus becomes
\begin{align}
    f(y_i|p)={n \choose y_i}\left (\frac{e^\eta}{1+e^\eta}\right )^{y_i} \left (1-\frac{e^\eta}{1+e^\eta} \right )^{n-y_i}
\end{align}
and the sample likelihood is again the product of the individual likelihood contributions of each observation. For details, see also \cite{Faraway2016} and \cite{McCullagh1989}. For a full Bayesian analysis, we need to specify priors $p(\alpha)$ and $p(\bm{\beta})$ on both the intercept $\alpha$ and the regression coefficient vector $\bm{\beta}$. In the following we use weakly informative priors, which are also the recommended default priors used in the \texttt{rstanarm} package \citep{Goodrich2020}. The \texttt{rstan} \citep{RStan2020} and \texttt{rstanarm} packages will be used for posterior inference via the No-U-Turn Hamiltonian Monte Carlo sampler of \cite{Hoffman2014} in Stan \citep{Carpenter2017}. Unless otherwise stated, we use $\alpha \sim \mathcal{N}(0,2.5)$ and $\beta_k \sim \mathcal{N}(0,2.5)$ for all $k=1,...,K$ regression coefficients \citep{Gabry2020RstanarmPriorsVignette}.

The posterior distribution is then obtained via MCMC sampling as
\begin{align}
    f(\alpha,\bm{\beta}|\bm{X},y)\propto p(\alpha)\times \prod_{k=1}^K p(\beta_k) \times \prod_{i=1}^n \left (\frac{e^{\eta_i}}{1+e^{\eta_i}}\right )^{y_i}\left (\frac{1}{1+e^{\eta_i}} \right )^{n-y_i}    
\end{align}
Four chains with $n=4000$ iterations are used for posterior inference in all models, where $n=2000$ iterations are used as a burn-in. Convergence to the posterior distribution is checked visually \citep{Gabry2019} and numerically using the Gelman-Rubin shrink factor $\hat{R}$ \citep{Gelman1992}.

\subsection*{Data set}
In the running example, we use data from the \textit{Western Collaborative Group Study} (WCGS) on chronic heart disease \citep{Rosenman1975}, see also \cite{Faraway2016}. The WCGS began in $1960$ with $n=3524$ male volunteers who were employed by $11$ California companies. Subjects were $39$ to $59$ years old and free of heart disease as determined by an electrocardiogram. After the initial screening, the study population dropped to $3154$ and the number of companies to $10$ because of various exclusions. At baseline the following information was collected: socio-demographic including age, education, marital status, income, occupation; physical and physiological measurements including height, weight, blood pressure, electrocardiogram, and corneal arcus; biochemical measurements including cholesterol and lipoprotein fractions; medical and family history and use of medications; behavioural data based on interviews, smoking, exercise, and alcohol use. Later surveys added data on anthropometry, triglycerides, Jenkins Activity Survey, and caffeine use. Average follow-up continued for 8.5 years with repeat examinations.

For the running example, a subset of $n=3140$ participants was selected based on the exclusion criterion of any missing values for a predictor. While the size of this dataset is not huge and for illustration purposes only a small subset of the available predictors for the outcome coronary heart disease (absent/present) is used, it is clear that precise LOO-CV means refitting any model under consideration $3140$ times. Even for a moderate-dimensional posterior distribution, this computational effort becomes massive, so that when comparing a moderate number of models, the benefits of approximate LOO-CV in particular with subsampling are substantial.
\begin{center}
\begin{table}[h!]
\normalsize\sf\centering
Models used for illustration purposes with the WCGS data set 
\begin{tabular}{p{1cm}p{10cm}}
\hline
Model & Predictors\\
$M_1$ & age, height, weight\\
$M_2$ & age, height, weight, systolic blood pressure, diastolic blood pressure, fasting serum cholesterol\\
$M_3$ & age, height, weight, systolic blood pressure, diastolic blood pressure, fasting serum cholesterol, number of cigarettes smoked per day, arcus senilis factor
\end{tabular}
\caption{Models used for illustration purposes with the WCGS data set}
\label{tab:models}
\end{table}
\end{center}
For illustration purposes, we use three competing models in the example, which are shown in table \ref{tab:models}. The first model $M_1$ includes the age in years, the weight in pounds and the height in inches of study participants. The second model $M_2$ additionally includes systolic and diastolic blood pressure in mm Hg of each participant, and the fasting serum cholesterol in mm \%. The third model $M_3$ additionally includes the number of cigarettes smoked per day as well as the \textit{arcus senilis} factor (absent or present). The arcus senilis factor is true if a depositing of phospholipid and cholesterol in the peripheral cornea in patients is visible which appears as an opaque ring.

\subsection*{Schematic execution of approximate PSIS-LOO-CV with subsampling}
To obtain the approximate PSIS-LOO-CV estimate of the expected log predictive density in equation (\ref{eq:elpdbar}), we follow these steps:
\begin{enumerate}
    \item{We approximate the posterior distribution via Laplace approximation in Stan \citep{Blei2017, Carpenter2017}.\footnote{This is done via the \texttt{optimising} function in the \texttt{rstan} package \citep{RStan2020}. Steps 2. to 5. can be carried out using the \texttt{loo} package in R \citep{VehtariLOOPackage2020}.}}
    \item{We approximate $\tilde{\pi}_i \propto \log p(y_i|y)$ for all $n$ observations using the posterior mean $\theta_L$ of the Laplace approximation as $\tilde{\pi}\propto - \log p(y_i|\theta_L)$. We assume that $n$ is large enough for the approximation to hold.}
    \item{We sample $m < n$ observations using probability-proportional-to-size subsampling and compute $\hat{p}(y_i|y_{-i})$ in equation (\ref{eq:importanceSamplingApproximation}) as an estimate of the leave-one-out log posterior $\log p_M(y_i|y_{-i})$ in equation (\ref{eq:elpd_LOO-CV}).} 
    \item{The Pareto $\hat{k}$ diagnostic values are used to diagnose if the Pareto-smoothed importance sampling of each term $\hat{p}(y_i|y_{-i})$ is reliable. For the problematic observations with $\hat{k}>.7$, precise calculation of $\log p_M(y_i|y_{-i})$ is conducted. The Pareto-$\hat{k}$ diagnostic values additionally assess the reliability of the posterior approximation: If there are too many observations $y_i$ with $\hat{k}>.7$, the adapted importance sampling scheme which uses the posterior approximation is not reliable \citep{Magnusson2019}.}
    \item{We calculate $\hat{\overline{\text{elpd}}}_{\text{LOO}}(M)$ in equation (\ref{eq:probpropEstLOOCV}) based on the results of step 3. We calculate the subsampling error $\mathbb{V}[\hat{\overline{\text{elpd}}}_{\text{LOO}}(M)]$ as given in equation (\ref{eq:subsamplingError}) to check if the precision based on $m$ subsamples is high enough. For model comparison, we calculate $\hat{\sigma}_{\text{LOO}}^2(M)$ as an estimate for $\mathbb{V}[\overline{\text{elpd}}_{\text{LOO}}(M)]$ as given in equation (\ref{eq:elpdErrorModelComparison}).}
\end{enumerate}
Notice that if the variance $\mathbb{V}[\hat{\overline{\text{elpd}}}_{\text{LOO}}(M)]$ is too large, repeating step 4 with increased $m$ and then running steps 5 and 6 again is possible until the variance is small enough. However, increasing subsample size $m$ in turn causes higher computational effort, so interest lies in using the smallest subsample size $m$ which yields a precise enough estimator $\hat{\overline{\text{elpd}}}_{\text{LOO}}(M)$. If $m=n$, then the subsampling procedure defaults to PSIS-LOO-CV which uses the full sample size $n$, and the only difference is that the posterior is obtained faster via Laplace approximation instead of precise MCMC inference.

In the data set at hand, approximating the posterior will speed up computations, but for illustration purposes, we first obtain the posterior distribution via full MCMC in step one. Later, we will use Laplace approximation instead and show that the differences in ELPD estimates are only subtle between both methods. Notice that the approximation method and the use of subsampling are not dependent on each other, that is one can easily use subsampling with full MCMC inference or approximate posterior inference (Laplace or variational inference) without subsampling.\footnote{However, using the posterior mean $\hat{\theta}$ of the Laplace approximation is much more efficient when probability-proportional-to-size subsampling is used compared to using the posterior mean $\hat{\theta}$ based on full MCMC inference to compute $\tilde{y_i} \propto -\log p(y_i|\hat{\theta})$.} In situations where obtaining a posterior distribution is not that costly (the number of model parameters $p$ is only moderate) and the sample size $n$ is huge (compare table \ref{tab:1}, lower left cell), using full MCMC inference with subsampling is recommended. In settings where the number of parameters $p$ is large, but sample size $n$ is small, using posterior approximations without subsampling is recommended (compare table \ref{tab:1}, upper right cell), and in situations in which both $p$ and $n$ are large (compare table \ref{tab:1}, lower right cell), using both approximate posterior inference and subsampling with $m<<n$ is recommended.\footnote{When both $n$ and $p$ are moderate, using full MCMC inference and no subsampling is always recommended, as no approximation error and no subsampling error is induced. However, in most realistic psychological models the number of parameters quickly becomes large (e.g. multilevel models), and in times of big data the sample size $n$ often is huge.}

\subsection*{Inference via precise MCMC and no subsampling}
In this section, all model posteriors were obtained with precise MCMC inference and no subsampling. The next section then shows how to use subsampling in combination with precise MCMC inference, of which precise MCMC inference without subsampling is just a special case when $m=n$. The section thereafter then shows how to combine subsampling with approximate posterior inference. Using only approximate posterior inference without subsampling again is just a special case of the above when $m=n$.

Models were fitted via the \texttt{rstanarm} and \texttt{rstan} packages \citep{Goodrich2020, RStan2020} using the logit model and priors as specified above. $K$-fold CV and precise LOO-CV were also conducted via the \texttt{rstanarm} package, and IS-LOO-CV, PSIS-LOO-CV and approximate PSIS-LOO-CV (with subsampling) were carried out using the \texttt{loo} package \citep{VehtariLOOPackage2020} in R \citep{RProgrammingLanugage}.
\begin{center}
\begin{table}[h!]
\normalsize\sf\centering
ELPD estimates of different model selection methods in the $\mathcal{M}$-open setting for the \textit{Western Collaborative Heart Study} \citep{Rosenman1975}\\
\begin{tabular}{llllll}
\hline
Model & LOO-CV & $10$-fold CV & IS-LOO-CV & PSIS-LOO-CV & WAIC \\
$M_1$ & -859.5 & -857.9 & -859.5 & -859.5 & -859.5\\
$M_2$ & -817.8 & -816.8 & -817.7 & -817.7 & -817.7\\
$M_3$ & -804.4 & -806.0 & -804.4 & -804.4 & -804.4
\end{tabular}
\caption{ELPD estimates of different model selection methods in the $\mathcal{M}$-open setting for the WCGS data set; all posteriors were obtained via precise MCMC inference}
\label{tab:results}
\end{table}
\end{center}
\begin{figure*}[h!]
    \centering
    \includegraphics[scale=0.75]{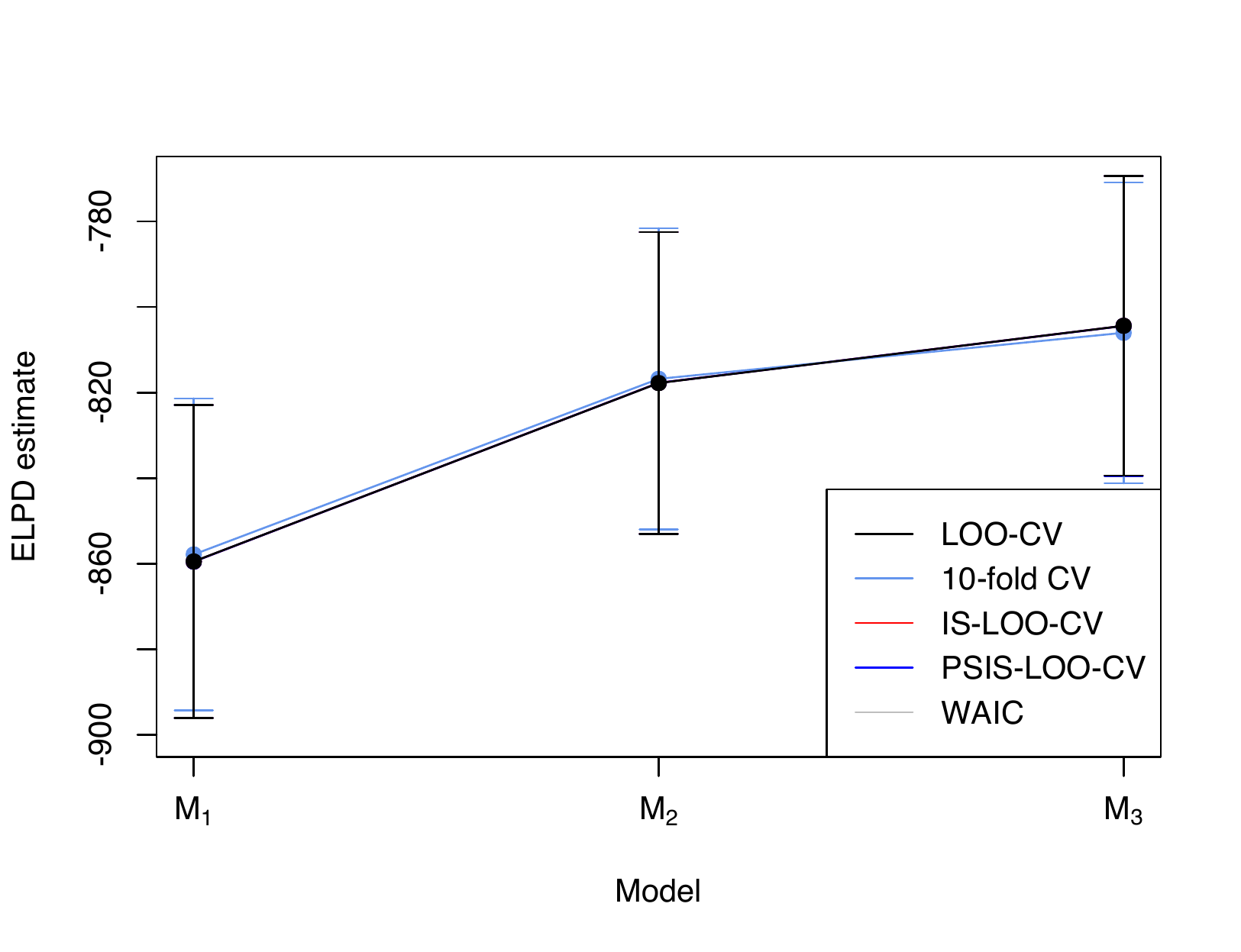}
    \caption{Model comparison and selection based on the expected log predictive density estimates using different model selection methods for models $M_1,M_2,M_3$ for the WCGS data set; all posterior distributions were obtained via full MCMC and no subsampling was used}
    \label{fig:ELPDestimates}
\end{figure*}
Table \ref{tab:results} and figure \ref{fig:ELPDestimates} show the results of the various model comparison techniques for the three models $M_1,M_2$ and $M_3$ given in table \ref{tab:models}. LOO-CV is the gold standard, as all other methods try to estimate $\overline{\text{elpd}}_{\text{LOO}}(M)$ while reducing the computational effort. To make comparisons easy, for all methods the expected log predictive density (ELPD) estimate is reported. For precise LOO-CV, this is given in equation (\ref{eq:elpd_LOO-CV}) (analogue for $K$-fold CV). For IS-LOO-CV equation (\ref{eq:importanceSamplingApproximation}) is used to estimate the terms in equation (\ref{eq:elpd_LOO-CV}), and for PSIS-LOO-CV the importance sampling ratios are adapted as described in section \ref{sec:psisloo}. The WAIC estimate is computed as given in equation (\ref{eq:waic}).

Based on the results in table \ref{tab:results}, model $M_3$ yields the best out-of-sample predictive ability. While precise LOO-CV is the gold standard, the computation for each model means refitting the logistic regression model $3140$ times, so that for the three models, $\approx 10000$ posteriors need to be fitted. On a regular machine, this takes about 15 hours of CPU time. In contrast, using $K$-fold CV already reduces the computation time substantially and estimating the ELPD only takes a few minutes for each model. IS-LOO-CV is even more efficient and takes only a few seconds, but can be highly unstable when the variance of the importance ratios becomes infinite (for details about the relationship between the magnitude of $\hat{k}$ and the variance of the importance ratios see \citep{Vehtari2015}). PSIS-LOO-CV solves this problem and is more reliable as it indicates problems with the importance ratios via the Pareto $\hat{k}$ diagnostic values. However, PSIS-LOO-CV also takes longer to compute (about 2 minutes for the WCGS data set for each model). While for this dataset, both $K$-fold CV and PSIS-LOO-CV only take moderate amounts of time to estimate the ELPD, for increasing sample size $n$ and number of parameters $p$ both methods become quickly inefficient. Note that WAIC is computationally less costly, but (1) computing the expected log predictive densities required for computation becomes costly for large $n$, too, and (2) based on the results of \cite{Piironen2017a, Vehtari2017}, WAIC is much less robust for model comparison than PSIS-LOO-CV.

Notice that WAIC, IS-LOO-CV and PSIS-LOO-CV yield identical results based on one-digit precision, and are extremely close to the precise LOO-CV estimates, compare figure \ref{fig:ELPDestimates}.

\subsection*{Model comparison and selection}
\begin{figure*}[h!]
    \centering
    \includegraphics[scale=0.8]{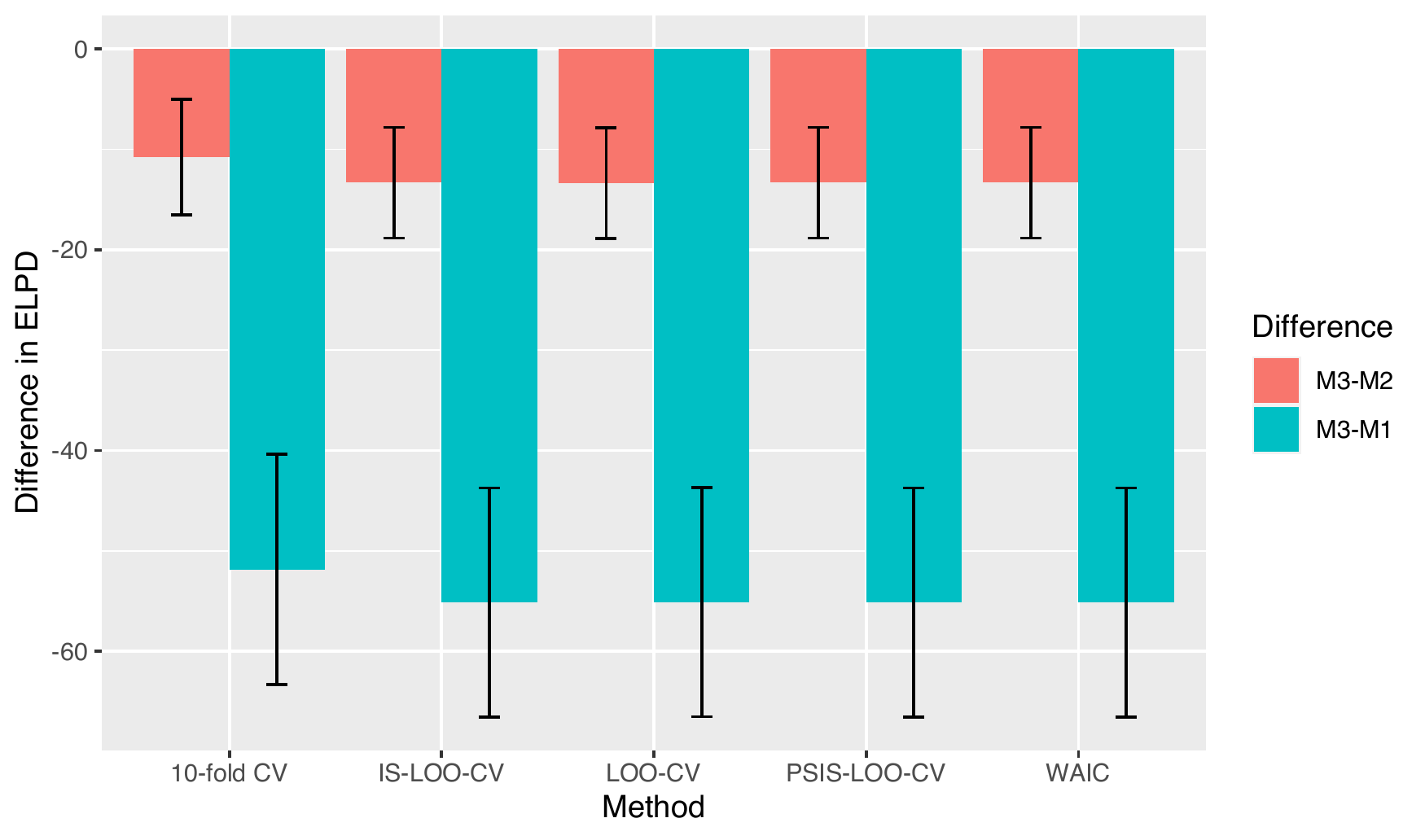}
    \caption{Model comparison and selection based on the expected log predictive density estimates using different model selection methods for models $M_1,M_2,M_3$ for the WCGS data set; all posterior distributions were obtained via full MCMC and no subsampling was used}
    \label{fig:barplot}
\end{figure*}

To compare and select among the competing models, the difference in ELPD is computed for each of the models using the corresponding ELPD estimators (LOO-CV, $10$-fold CV, IS-LOO-CV, PSIS-LOO-CV or WAIC). Standard errors are obtained as the empirical standard deviation from the posterior draws of differences in ELPD subsequently, and figure \ref{fig:barplot} shows the differences in ELPD between models $M_3$ and $M_1$ (M3-M1) and $M_3$ and $M_2$ (M3-M2) as well as the standard errors of the ELPD differences. Based on the results, it becomes clear that all methods provide similar results, and the error bars indicate that model $M_3$ has a larger expected log predictive density (and thereby a better predictive ability) than model $M_2$ or model $M_1$ after taking the standard errors into account. While all methods show similar results and favour model $M_3$, notice that the computational effort to obtain the ELPD estimate differs substantially between each of the techniques.

\subsection*{Investigating the Pareto $\hat{k}$-values}
\begin{figure*}[h!]
    \begin{subfigure}[b]{0.49\textwidth}
            \centering
            \includegraphics[width=\textwidth]{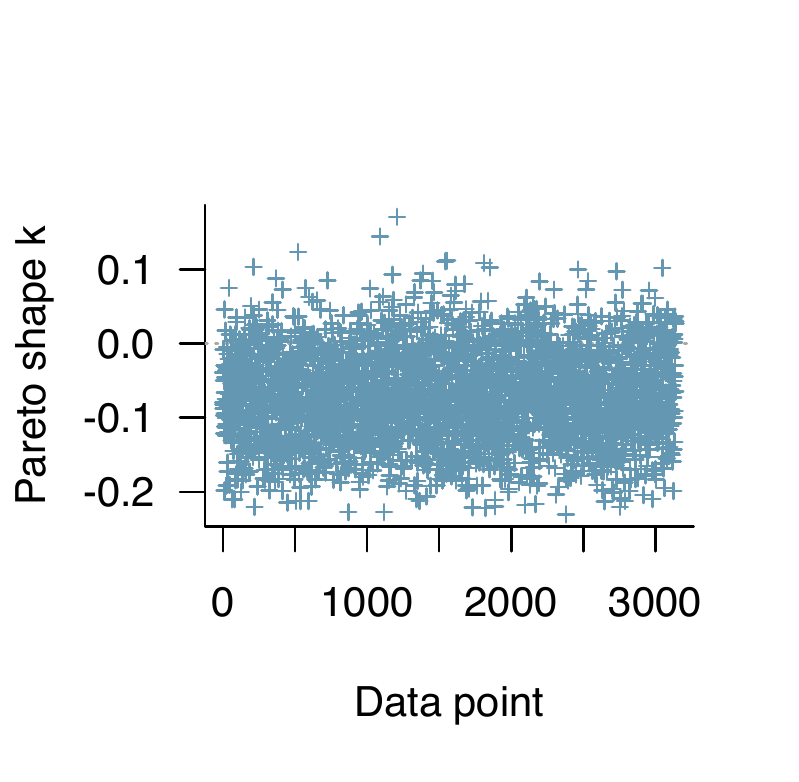}
                \caption{Pareto $\hat{k}$ diagnostic plot for model $M_1$}
    \label{fig:paretokdiagnosticsM1}
    \end{subfigure}
    \begin{subfigure}[b]{0.49\textwidth}
            \centering
            \includegraphics[width=\textwidth]{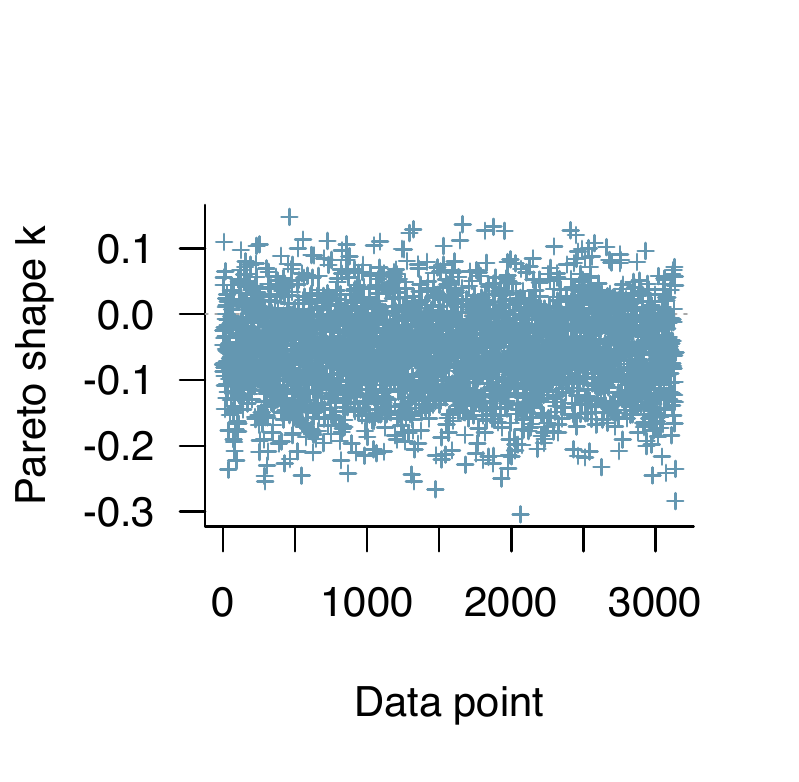}
            \caption{Pareto $\hat{k}$ diagnostic plot for model $M_2$}
    \label{fig:paretokdiagnosticsM2}
    \end{subfigure}
    \centering
    \begin{subfigure}[b]{0.49\textwidth}
            \centering
            \includegraphics[width=\textwidth]{paretokdiagnosticM2.pdf}
            \caption{Pareto $\hat{k}$ diagnostic plot for model $M_3$}
    \label{fig:paretokdiagnosticsM2}
    \end{subfigure}
    \caption{Pareto $\hat{k}$ diagnostic plot for models $M_1$, $M_2$ and $M_3$}
    \label{fig:paretokdiagnosticplots}
\end{figure*}

Investigating the Pareto $\hat{k}$-values for diagnosing if PSIS-LOO-CV works as expected is mandatory when using the method. Figure \ref{fig:paretokdiagnosticplots} shows the corresponding Pareto $\hat{k}$ diagnostic plots produced via the \texttt{loo} package \citep{VehtariLOOPackage2020}. For each observation $y_i$, the corresponding $\hat{k}$ value used in the generalised Pareto distribution which is fit to the largest importance ratios $w(\theta_s)$ is shown, and values with $\hat{k}>.7$ indicate that the variance of the importance ratios $w(\theta_s)$ may be too large for PSIS-LOO-CV to work reliably. If the number of observations for which $\hat{k}>.7$ is moderate, the corresponding LOO posteriors can be refitted manually without causing a substantial increase in the computation time. However, if the number of observations with problematic Pareto $\hat{k}$-values becomes large, both the reliability of the importance sampling scheme and the computational efficiency of the method are lost and $K$-fold CV may be more efficient. As can be seen from the plots in figure \ref{fig:paretokdiagnosticplots}, all $\hat{k}$-values are smaller than $0.7$ for all three models, indicating that the variance of the importance ratios $w(\theta_s)$ is not inflated and the method worked reliably. However, this does \textit{not} imply that the model for which the ELPD estimate is obtained via PSIS-LOO-CV has a good out-of-sample predictive ability. It only ensures the proper working of the PSIS-LOO-CV method itself.

\subsection*{Inference via precise MCMC with subsampling}
\begin{figure*}[h!]
    \centering
    \includegraphics[scale=0.7]{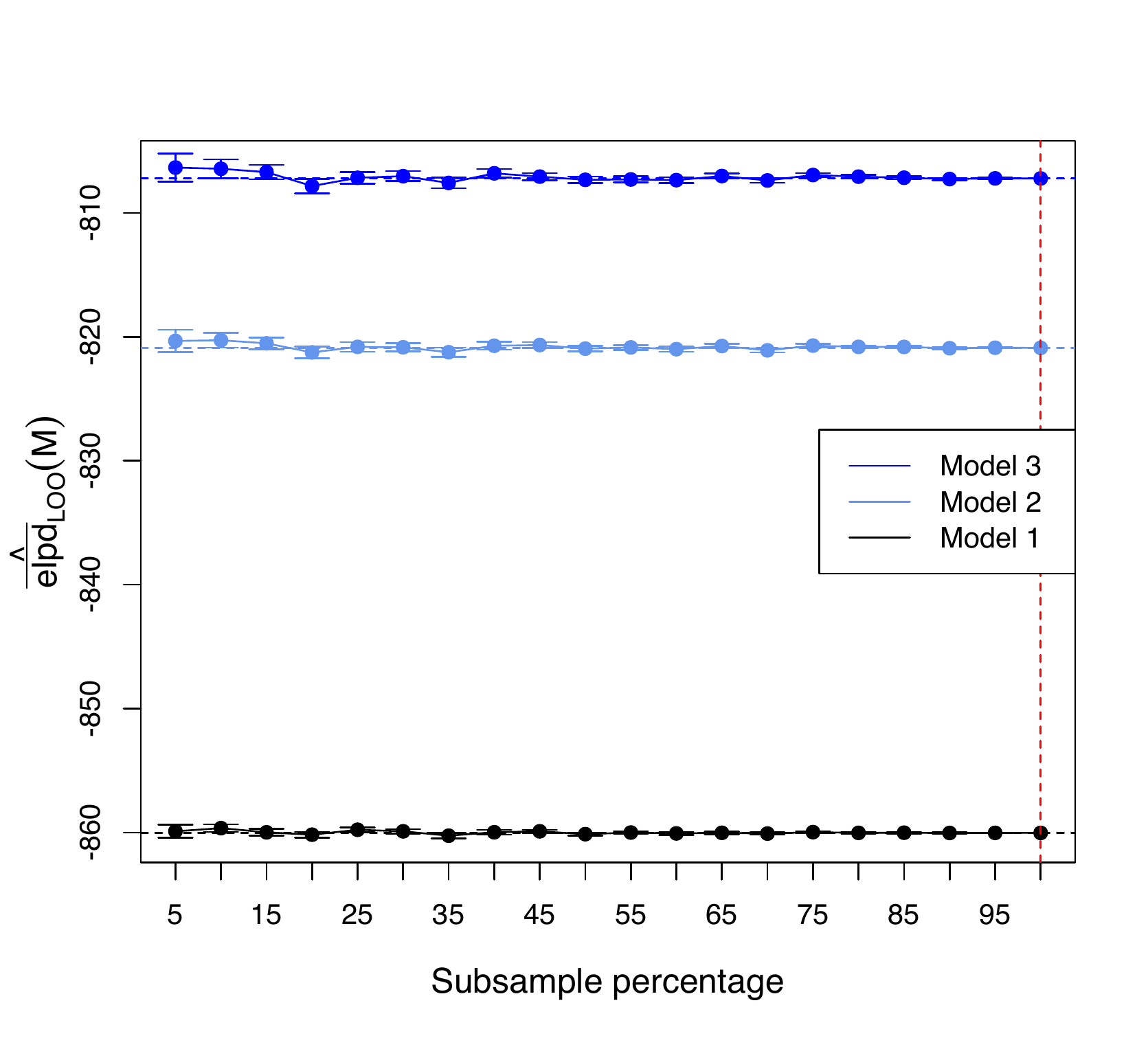}
    \caption{Model comparison and selection based on the expected log predictive density estimates using different model selection methods for models $M_1, M_2, M_3$ for the WCGS data set; all posterior distributions were obtained via full MCMC and subsample sizes range from $5\%$ to $100\%$ of the sample size $n$; horizontal dashed lines are ELPD estimates based on PSIS-LOO-CV using the full sample. Points on the vertical dashed line are estimates based on the full sample size.}
    \label{fig:subsamplingMCMC}
\end{figure*}

Figure \ref{fig:subsamplingMCMC} shows the results of refitting the models, and this time ELPD estimates were obtained via PSIS-LOO-CV with subsampling. On the $x$-axis, subsample percentages ranging from $5\%$ to $100\%$ are given, where $100\%$ means $m=n$ so that the methods becomes PSIS-LOO-CV and has no computational advantages anymore. The horizontal dashed lines are the corresponding PSIS-LOO-CV estimates given in table \ref{tab:results}, and for increasing $m$ the subsample ELPD estimates $\hat{\overline{\text{elpd}}}_{\text{LOO}}(M)$ estimates approach the PSIS-LOO-CV estimate as expected.

From figure \ref{fig:subsamplingMCMC} it becomes clear that even when only a very small fraction of the original samples are used, the ELPD estimates based only on a subsample of size $m$ of the original full sample of size $n$ are already quite accurate. For example, using 5 per cent of the sample size for subsampling (that is, only $m=n/20=3140/20=157$ samples), the ELPD estimates are already significantly different for the three models $M_1, M_2, M_3$, and the subsampling error bars already include the PSIS-LOO-CV estimate which is based on the full sample size.

\begin{figure*}[h!]
    \begin{subfigure}[b]{0.49\textwidth}
            \centering
            \includegraphics[width=\textwidth]{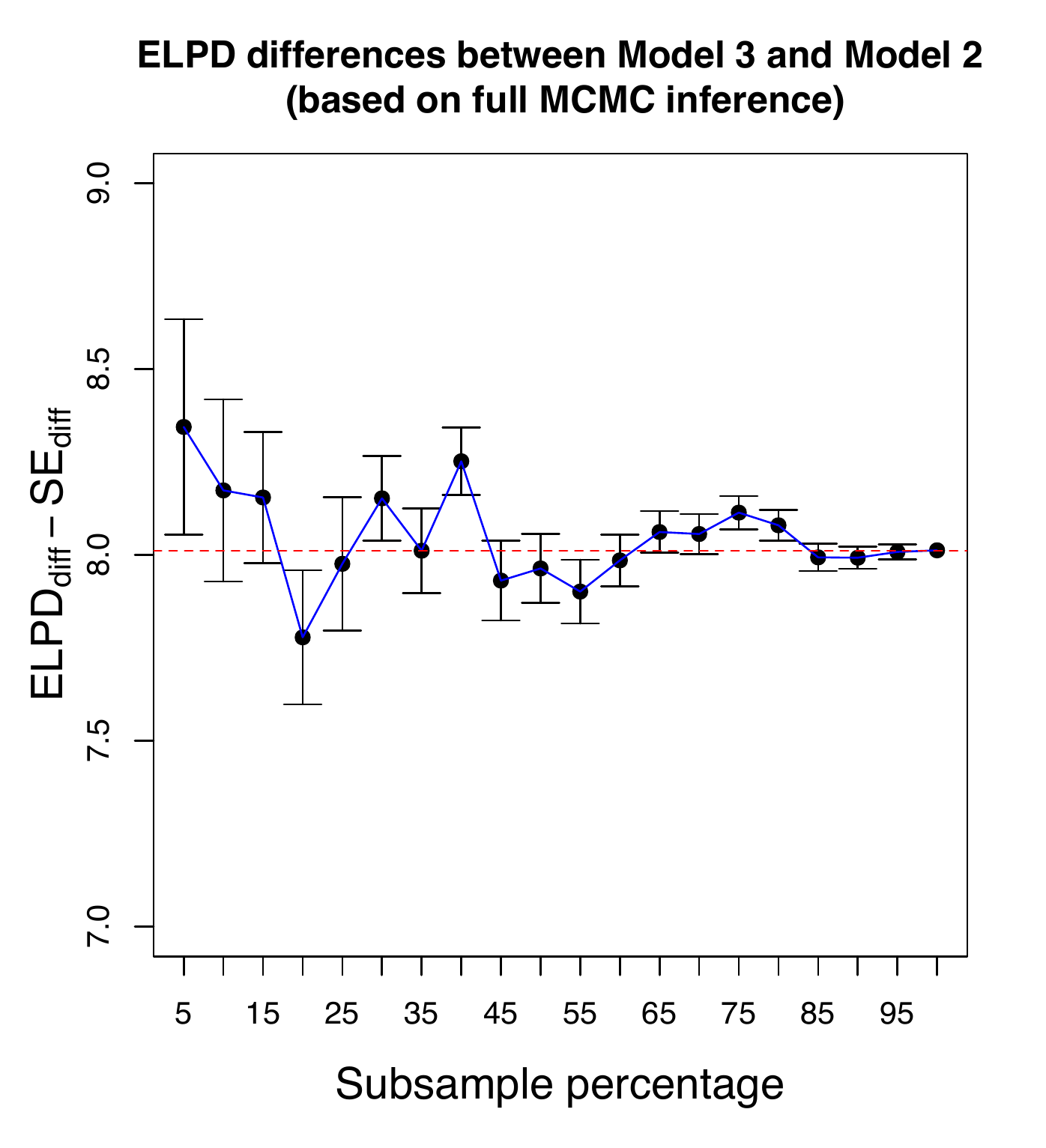}
                \caption{ELPD differences between model $M_3$ and $M_2$ based on full MCMC inference and subsampling}
    \label{fig:ELPDdiffM3M2}
    \end{subfigure}
    \begin{subfigure}[b]{0.49\textwidth}
            \centering
            \includegraphics[width=\textwidth]{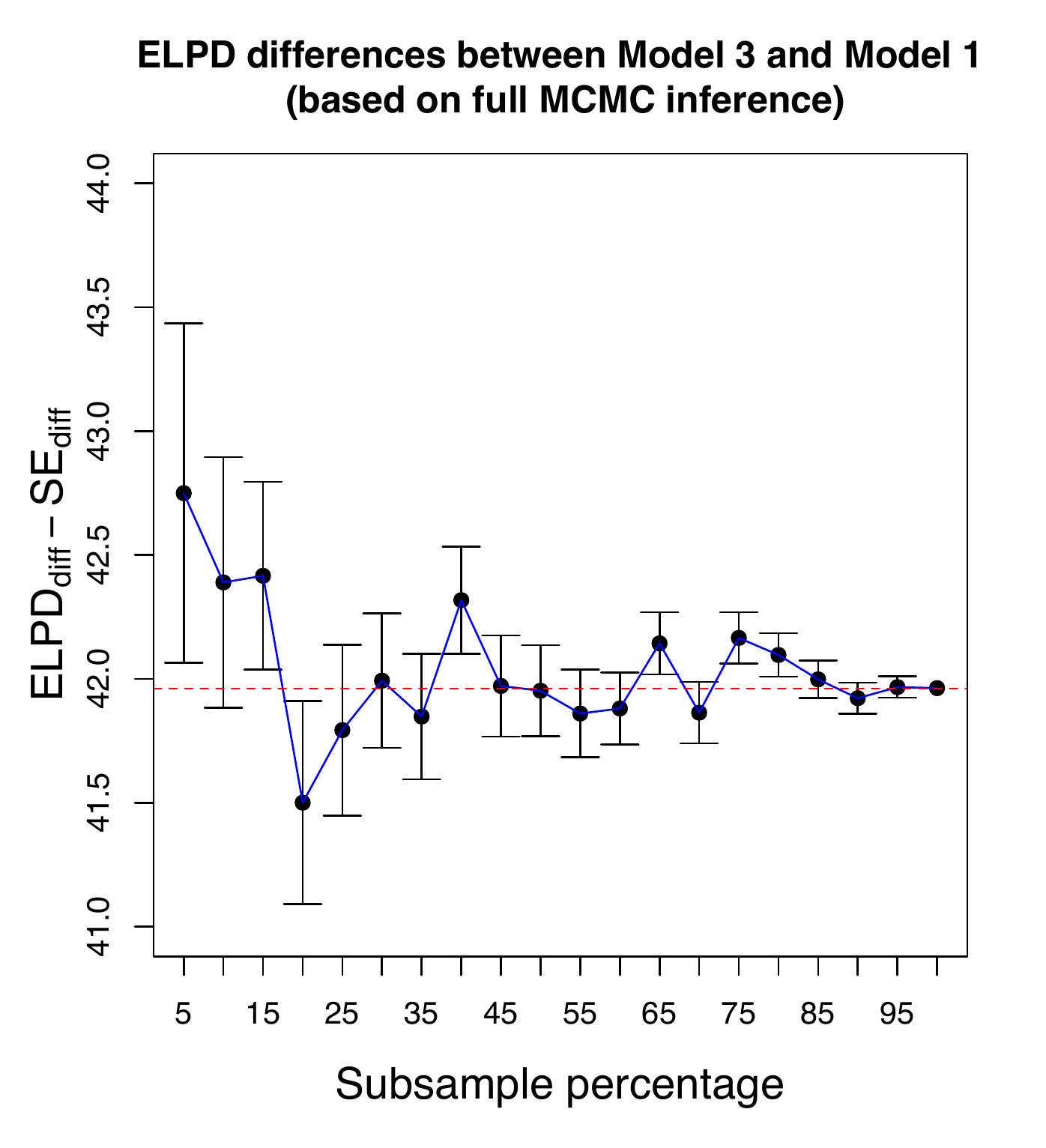}
            \caption{ELPD differences between model $M_3$ and $M_1$ based on full MCMC inference and subsampling}
    \label{fig:ELPDdiffM3M1}
    \end{subfigure}
    \caption{ELPD differences between models $M_3$ and $M_1$ and $M_3$ and $M_2$; Error bars show the subsampling error, which decreases for increasing subsample size $m$; Posteriors are based on full MCMC inference and subsamples used are identical for each subsample percentage for all models}
    \label{fig:ELPDdiffMCMC}
\end{figure*}

Figure \ref{fig:ELPDdiffMCMC} shows the relationship between ELPD differences and subsample sizes used. The ELPD differences between models $M_3$ and $M_1$ and between models $M_3$ and $M_2$ are shown. The standard error $SE_{\text{diff}}$ has been computed from equation (\ref{eq:elpdErrorModelComparison}), and the error bars show the subsampling error as given in equation (\ref{eq:subsamplingError}), which decreases to zero. Taking into account both the standard error and subsampling error, even for $m=n/10=314$ subsamples the difference in expected log predictive density between models $M_3$ and $M_2$ is $\approx 8.2$, and between models $M_3$ and $M_1$ is $\approx 42.4$, indicating that $M_3$ yields the best predictive ability of the three models (compare also figure \ref{fig:subsamplingMCMC}). Notice that in both cases, the subsampling error bars include the PSIS-LOO-CV value based on the full sample (the dashed red line in figures \ref{fig:ELPDdiffM3M2} and \ref{fig:ELPDdiffM3M1}). This indicates that even for about $m=10\%$, the ELPD estimates between PSIS-LOO-CV with subsampling and PSIS-LOO-CV based on the full sample are highly similar. However, while conclusions remain identical to the conclusions based on the ELPD when the full sample is used, the computational effort was reduced substantially by using such a small subsample. Notice also that in all subsample computations, identical subsamples were used in all steps for the three models, because otherwise the calculated expected log predictive density estimates may vary substantially depending on which subsamples have been selected in the subsampling procedure.

\subsection*{Inference via Laplace approximation and subsampling}
By now, precise MCMC inference was used for obtaining the posterior of all models. Using subsampling in combination with PSIS-LOO-CV significantly improved the computation time already while the posterior was still obtained via full MCMC. However, in large-dimensional posterior distributions, the differences in computation time between precise MCMC and approximate posterior inference become substantial. Therefore, in this section, we use Laplace approximation instead of precise posterior inference via MCMC in combination with subsampling. This reduces computation times even when the sample size $n$ and the number of predictors $p$ are large.

\begin{figure*}[h!]
    \centering
    \includegraphics[scale=0.7]{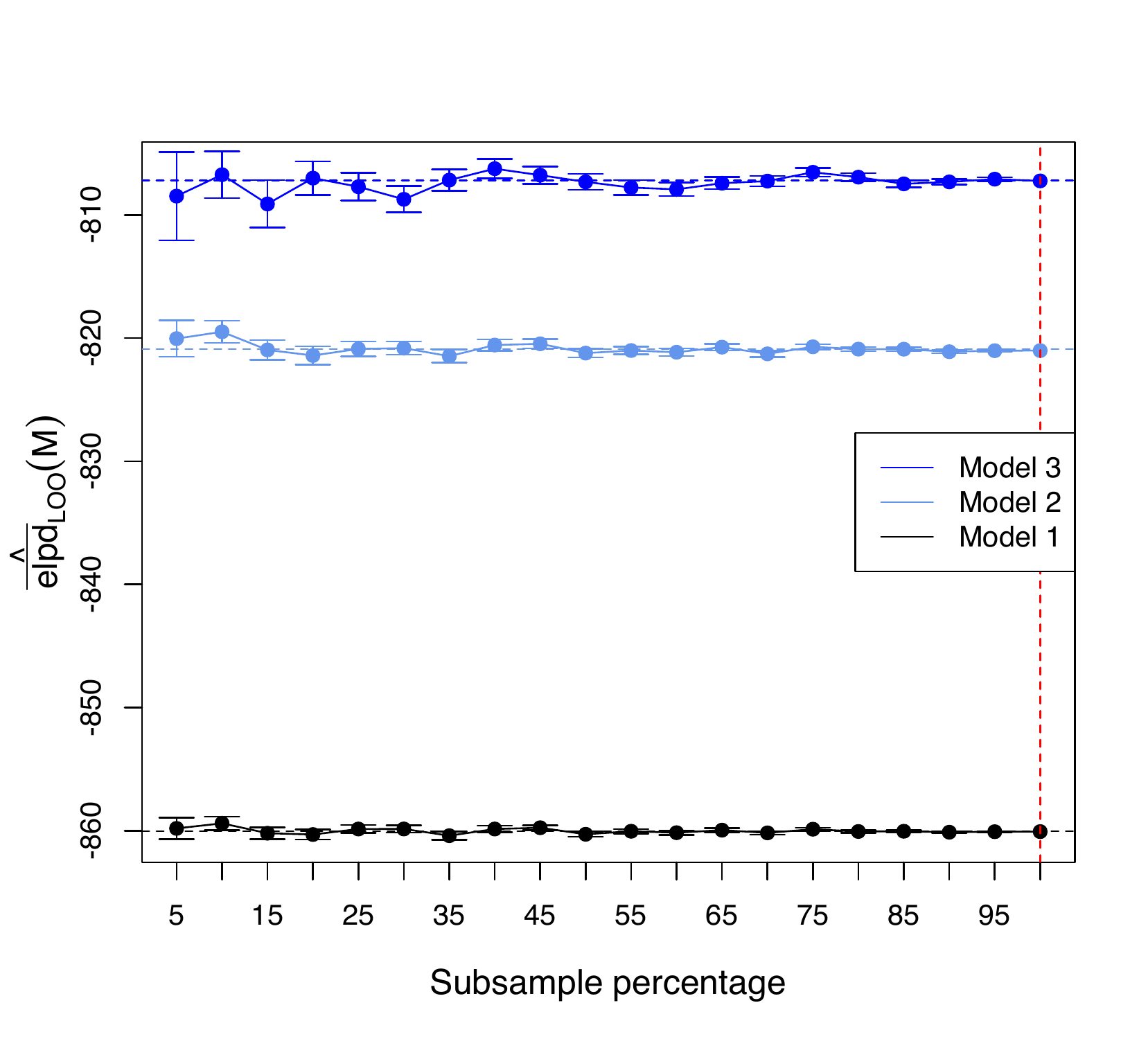}
    \caption{Model comparison and selection based on the expected log predictive density estimates using different model selection methods for models $M_1,M_2,M_3$ for the WCGS data set; all posterior distributions were obtained via Laplace approximation and subsample sizes range from $5\%$ to $100\%$ of the sample size $n$; horizontal dashed lines are the ELPD estimates based on PSIS-LOO-CV and were obtained via precise MCMC inference using the full sample; Points on the vertical dashed line are estimates based on the full sample size.}
    \label{fig:subsamplingLaplace}
\end{figure*}

Figure \ref{fig:subsamplingLaplace} shows the results when using subsampling in combination with Laplace approximation for posterior inference instead of precise MCMC inference. The differences in the ELPD estimates between full MCMC and the Laplace approximation are subtle, compare figures \ref{fig:subsamplingMCMC} and \ref{fig:subsamplingLaplace}. Based on the Laplace approximation, using $m=314$ subsamples the same conclusions are drawn as when using full MCMC inference and using $m=314$ subsamples. Notice that the conclusion drawn also is the same when using Laplace approximation and $m=314$ subsamples in contrast to full MCMC inference using all $n=3140$ samples.
\begin{figure*}[h!]
    \begin{subfigure}[b]{0.49\textwidth}
            \centering
            \includegraphics[width=\textwidth]{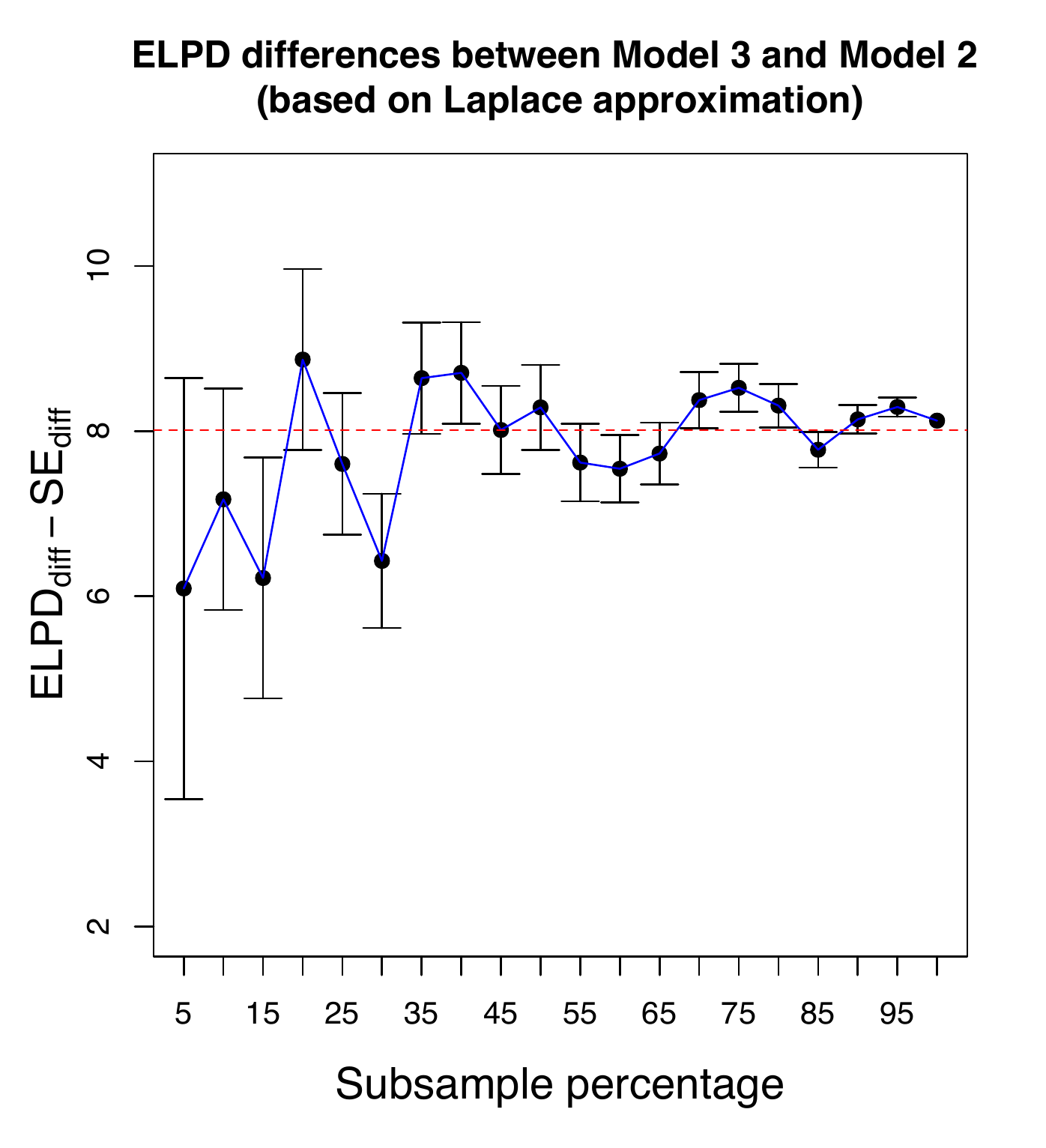}
                \caption{ELPD differences between model $M_3$ and $M_2$ based on Laplace approximations and subsampling}
    \label{fig:ELPDdiffM3M2Laplace}
    \end{subfigure}
    \begin{subfigure}[b]{0.49\textwidth}
            \centering
            \includegraphics[width=\textwidth]{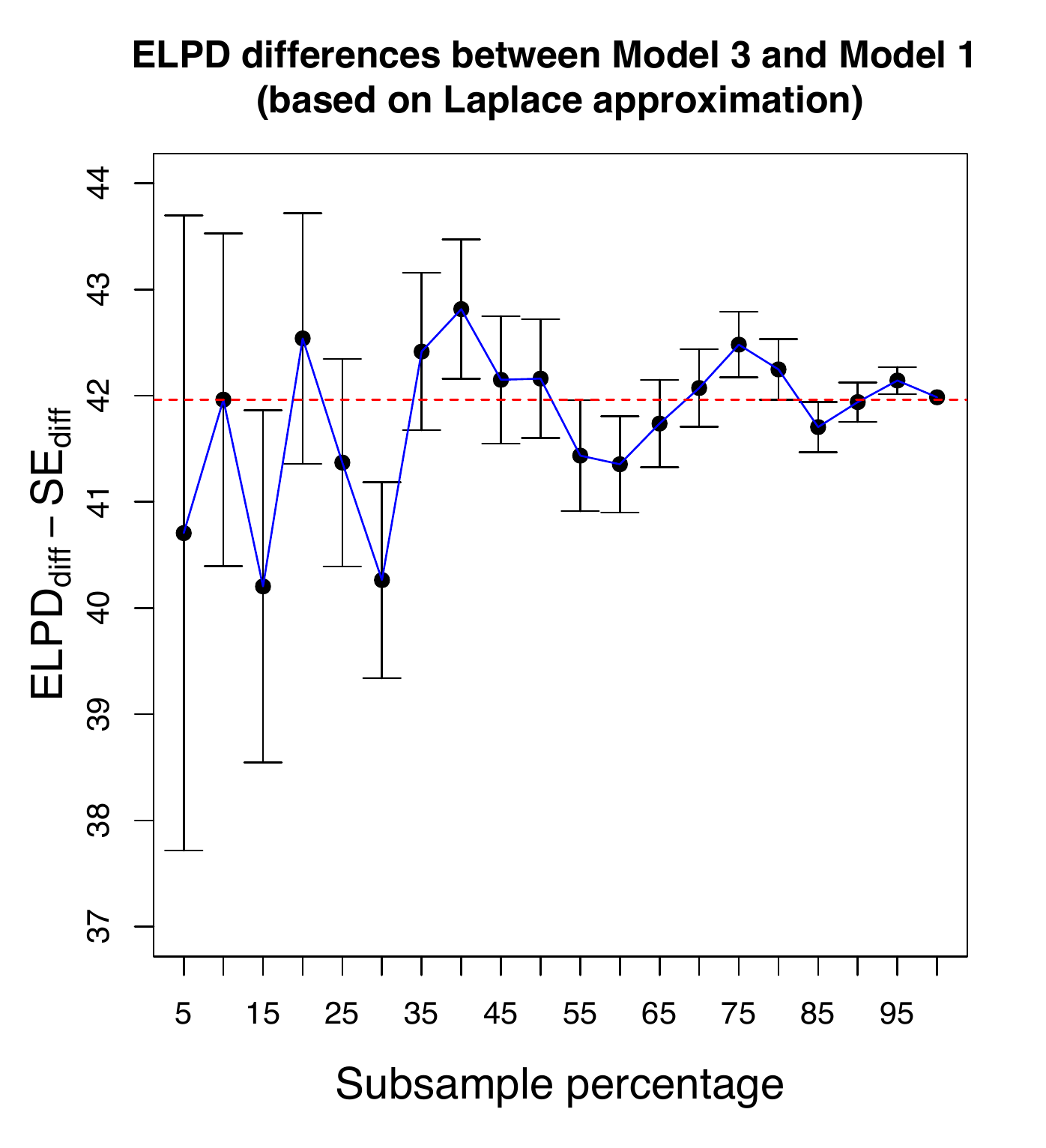}
            \caption{ELPD differences between model $M_3$ and $M_2$ based on Laplace approximations and subsampling}
    \label{fig:ELPDdiffM3M1Laplace}
    \end{subfigure}
    \caption{ELPD differences between models $M_3$ and $M_1$ and $M_3$ and $M_2$; Error bars show the subsampling error, which decreases for increasing subsample size $m$; Posteriors are based on Laplace approximation and subsamples used are identical for each subsample percentage for all models}
    \label{fig:ELPDdiffLaplace}
\end{figure*}

These results are also confirmed when analysing figure \ref{fig:ELPDdiffM3M2Laplace} and \ref{fig:ELPDdiffM3M1Laplace}, which again shows the normalised differences in estimated ELPD between models $M_3$ and $M_2$ and $M_3$ and $M_1$ when using Laplace approximation in combination with subsampling. For $m=n/10=314$, the difference in ELPD between models $M_3$ and $M_2$ is $\approx 7.2$, and between models $M_3$ and $M_1$ is $\approx 42$. For full MCMC inference, the corresponding differences in ELPD between models $M_3$ and $M_2$ were $\approx 8.2$, and between models $M_3$ and $M_1$ were $\approx 42.4$. Notice again, that the PSIS-LOO-CV estimate based on the full sample size $n$ and precise MCMC inference (dashed red line in figures \ref{fig:ELPDdiffM3M2Laplace} and \ref{fig:ELPDdiffM3M1Laplace}) is within the subsample error bars of the subsampled ELPD estimate when using $m=n/10=314$ subsamples for both model differences. The conclusions drawn are identical, but the reduction of the computational effort when using Laplace approximation instead of full MCMC is substantial. Notice that the difference between MCMC and Laplace approximation in ELPD estimates based on full sample size is given by the distance between the black point at $m=100\%$ and the height of the red dashed line. This difference is quite small, indicating that the Laplace approximation worked well.

\subsection*{Posterior predictive checks using probability integral transformations}
The preceding section showed how Bayesian model comparison and selection can be substantially fastened by using both approximate posterior inference in combination with probability-proportional-to-size subsampling. However, as mentioned above, although the $\hat{k}$-values signal no problems with the importance sampling ratios for the three models $M_1, M_2, M_3$, this does not necessarily indicate that \textit{any} model has a good predictive ability \citep{McElreath2020}. The ELPD is a quantitative estimate of a model's predictive ability, but the fact that model $M_3$ has a larger ELPD estimate does only imply that among the three models $M_1$, $M_2$ and $M_3$, model $M_3$ yields the best out-of-sample predictions. The predictive ability of all models can still be bad, which we show in the following. It is, therefore, crucial to perform marginal posterior predictive checks additionally in a fully Bayesian workflow when comparing competing models based on the expected log predictive density \citep{Gabry2019, Gelman2013BayesianDataAnalysis}. In addition to the sole comparison of the ELPD estimates and their differences, such marginal posterior predictive checks allow investigating if the models capture enough complexity or not.

The idea behind posterior predictive checks is simple: If a model is a good fit, it should
be able to generate data that resemble the actually observed data. To generate data for posterior predictive checks one usually simulates from the posterior predictive distribution 
\begin{align}
    p_M(\tilde{y}|y)=\int p_M(\tilde{y}|\theta)p_M(\theta|y)d\theta    
\end{align}
However, such posterior predictive checks (PPCs) make use of the observed data $y$ twice: Once for fitting the model $M$ and obtaining the posterior $p_M(\theta|y)$, and once for checking the posterior predictive fit based on $p_M(\tilde{y}|y)$. This can influence the ability of posterior predictive checks to detect problems between a model and its ability to predict new data $\tilde{y}$, in particular, if the model parameter and the posterior predictive check are related to each other. For example, comparing the mean of the observed data with the mean of the posterior predictive $p(\tilde{y}|y)$ when the model was a Gaussian model with unknown location and known standard deviation, the posterior mean parameter $\theta_M=\mathbb{E}[\theta|y]$ is influenced by the observed data $y$. Therefore, the mean of the posterior predictive $p(\tilde{y}|y)$ distribution will also be close to the posterior mean parameter $\theta_M$ and it will be difficult to detect differences between the posterior predictive and the observed data $y$.

\begin{figure*}[h!]
    \begin{subfigure}[b]{0.49\textwidth}
            \centering
            \includegraphics[width=\textwidth]{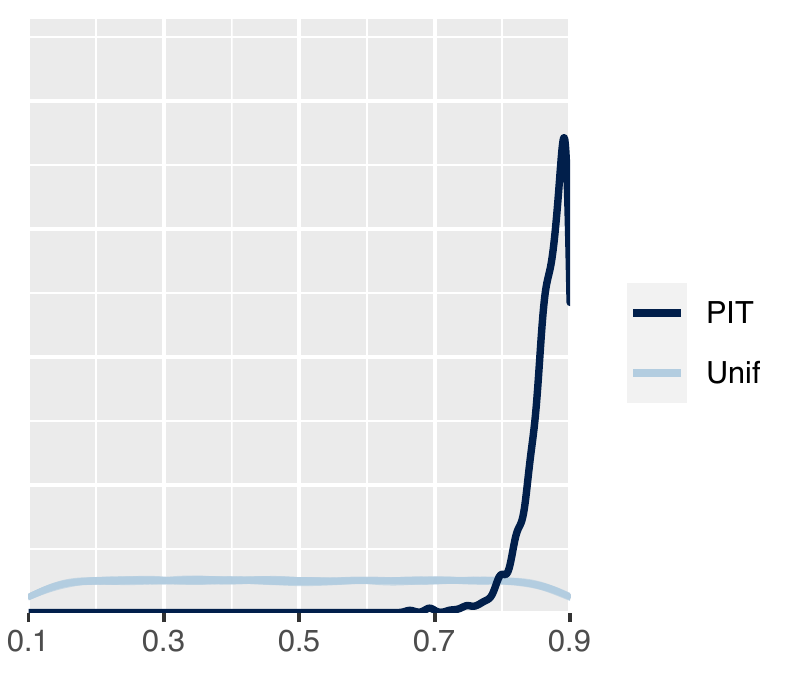}
                \caption{Marginal posterior predictive check for model $M_1$}
    \label{fig:loopitM1}
    \end{subfigure}
    \begin{subfigure}[b]{0.49\textwidth}
            \centering
            \includegraphics[width=\textwidth]{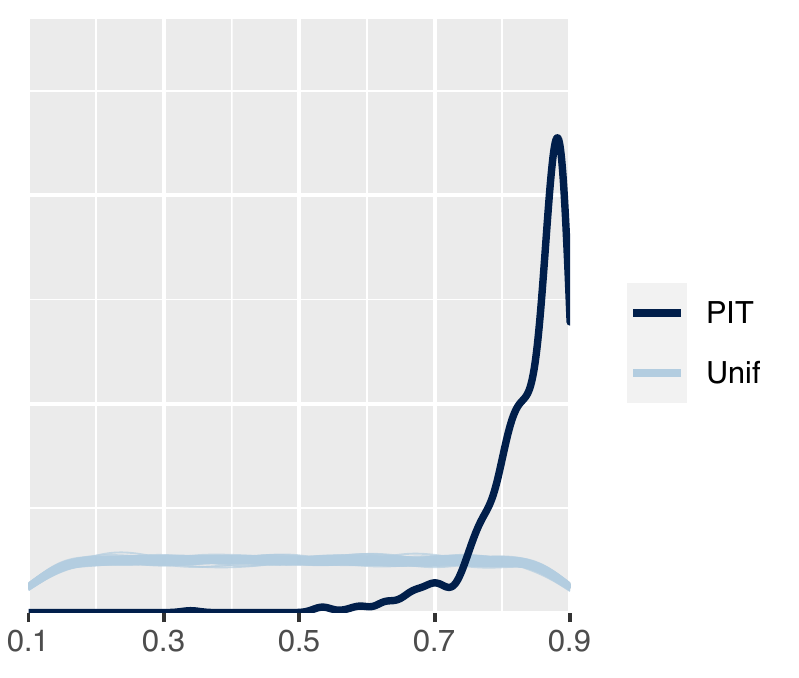}
            \caption{Marginal posterior predictive check for model $M_2$}
    \label{fig:loopitM2}
    \end{subfigure}
    \centering
    \begin{subfigure}[b]{0.49\textwidth}
            \centering
            \includegraphics[width=\textwidth]{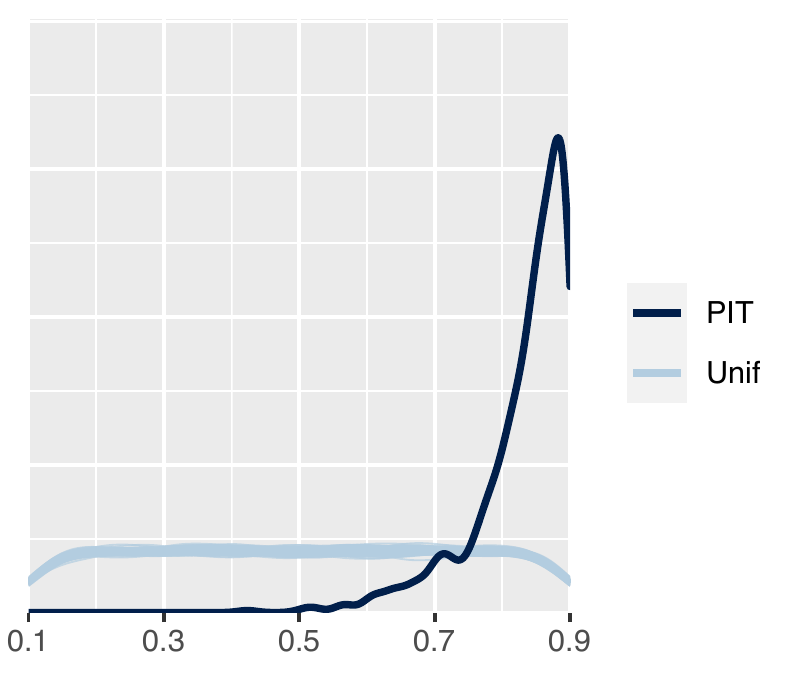}
            \caption{Marginal posterior predictive check for model $M_3$}
    \label{fig:loopitM3}
    \end{subfigure}
    \caption{Marginal posterior predictive checks for models $M_1$, $M_2$ and $M_3$}
    \label{fig:loopit}
\end{figure*}

While \cite{Gabry2019} proposes to use orthogonal statistics to the model parameters to evaluate the posterior predictive $p(\tilde{y}|y)$, a better solution is to use leave-one-out cross-validation posterior predictive distributions $p_M(y_i|y_{-i})$ instead for posterior predictive checks. In practice, one looks at the LOO-CV predictive cumulative density
function values, which are asymptotically uniform (for continuous data) when the model is
calibrated (for details see \cite{Gelfand1992} and \cite{Gelman2013BayesianDataAnalysis}). If a model $M$ is a good fit, the leave-one-out predictive values $p_M(y_i|y_{-i})$ should be equal for each observation $y_i$ because then the model does not predict some observations better or worse depending on which observation $y_i$ has been left out. Phrased differently, when the model predicts all observations $y_i$ equally well based on the LOO posterior predictive distributions $p_M(y_i|y_{-i})$, the model is not surprised by any observation $y_i$ and the distribution of the values $p_M(y_i|y_{-i})$ should be uniform. Marginal posterior predictive checks can be conducted easily using these so-called LOO-PIT values, and additionally, data is used only once compared to traditional PPCs. If the model is a good fit, the distribution of these LOO-PIT values should be uniform \citep{Gabry2019, Gelman2013BayesianDataAnalysis, Gelfand1992}. A simple but powerful method is therefore to compare the smoothed density of the LOO-PIT values with many independently generated samples from the uniform distribution. Figure \ref{fig:loopit} shows these marginal posterior predictive checks for the models $M_1$, $M_2$, and $M_3$, where the thick black line is the smoothed density estimate of the LOO-PIT values and the thin lines are smoothed densities of many independently generated random samples of the uniform distribution.

Although model $M_3$ has a slightly better (that is, less peaked) distribution than model $M_1$ and $M_2$, all three models are inadequate based on the results of the marginal posterior predictive checks. The reason is simple: All models are too simple, and they do not capture enough complexity to explain the risk of coronary heart disease based on such a small number of predictors (note that only a small subset of the predictors actually recorded in the study was used in each model). Incorporating additional relevant predictors will improve the PPC. Notice that such situations are to be expected often in the assumed $\mathcal{M}$-open view, and relying only on ELPD estimates is therefore not guarding against misinterpretations. However, the ELPD estimates provide a concise way to decide between a moderate number of competing models. Additionally conducted PPCs then provide a safeguard against overconfidence in one of the models, and guarantee that not all models under consideration are bad descriptions of the underlying data generating process $p(\tilde{y})$ (compare equation (\ref{eq:elpdbar})). 

Even when all models are bad descriptions, adding more predictors (and thereby also increasing the complexity of the model) may yield better PPCs, and this is when approximating the posterior via Laplace (or variational) inference and probability-proportional-to-size subsampling become highly efficient in contrast to traditional full MCMC inference based on the full sample size $n$. Estimating the ELPD even in large-scale models becomes therefore feasible by employing the approximate leave-one-out cross-validation methods discussed.

\section{Discussion}\label{sec:discussion}
Comparison of competing statistical models is an essential part of psychological research. From a Bayesian perspective, various approaches to model comparison and selection have been proposed in the literature. However, the model view strongly influences the applicability of these approaches in practice. This paper provided a tutorial on approximate leave-one-out cross-validation methods for Bayesian model comparison and selection in the $\mathcal{M}$-open view based on the expected log predictive density.

First, several perspectives on the model space $\mathcal{M}$ were discussed and the corresponding Bayesian model comparison methods in each view were analysed. Based on this discussion, the $\mathcal{M}$-open view seems to be the least restrictive and most realistic model view for a variety of psychological research. Often, no model under consideration is considered to be true, and there is also no reference model available.

Traditional methods in the $\mathcal{M}$-open view are various leave-one-out cross-validation (LOO-CV) techniques and a variety of information criteria. However, while these methods estimate the expected log predictive density of a model to judge the model's out-of-sample predictive ability, they become computationally inefficient when sample size $n$ or the number of model parameters $p$ becomes large. Therefore, this paper showed how to apply approximate LOO-CV, which is a recently developed computationally much more efficient method for Bayesian model comparison.

The Bayesian logistic regression model was used as a running example to compare three competing models using data from the \textit{Western Collaborative Group Study} on chronic heart disease \citep{Rosenman1975, Faraway2016}. Precise LOO-CV and WAIC were compared with standard importance sampling LOO-CV and Pareto-smoothed importance sampling LOO-CV, and both importance sampling-based approaches already improve upon LOO-CV and WAIC. While recent research has shown that WAIC is less robust for model selection than PSIS-LOO-CV (see \cite{Vehtari2017} for details; we omit these due to space reasons here), PSIS-LOO-CV does not scale when sample size $n$ and the number of predictors $p$ becomes large. Using posterior approximations in combination with probability-proportional-to-size subsampling as proposed by \cite{Magnusson2019}, we showed how computations can be sped up substantially. It was shown that even small subsample sizes $m$ of $\approx 10\%$ suffice to yield identical conclusions to full sample inference. Additionally, we showcased how Laplace approximation can be used to approximate the full posterior distribution and correct the importance sampling scheme for using a posterior approximation instead of full MCMC inference. This fastens the computation of the ELPD estimate even further. Here too, it was shown that already small subsample sizes $m$ of about $10\%$ of the original sample size $n$ suffice to yield identical conclusions to full sample Laplace inference, and even full sample MCMC inference.

Additionally, we showed how to perform marginal posterior predictive checks to safeguard against solely relying on the ELPD estimates provided by the method. These are substantial to prevent choosing the best model among a set of models which are all bad descriptions of the true data generating process.

It should be noticed that the method also has some limitations: First, posterior approximations may encounter trouble in approximating large-dimensional model posteriors, too. This is true in particular when predictors are correlated, an assumption which is often true in practical research. For example, \cite{Magnusson2019} showed that variational inference behaves badly already when $p=100$ and predictors are moderately correlated. However, the method provides convenient feedback in form of the Pareto $\hat{k}$ diagnostic values and marginal posterior predictive checks to indicate if the reliability of the provided ELPD estimates needs to be doubted. Second, the assumptions for the convergence of the estimator $\hat{\overline{\text{elpd}}}_{\text{LOO}}(M)$ to $\overline{\text{elpd}}_{\text{LOO}}(M)$ need to hold. However, these are not very restrictive (for details, see \cite{Magnusson2019}). 

Together, using probability-proportional-to-size subsampling and approximate posterior inference, a variety of statistical models frequently used in psychological research can be compared via estimation of the expected log predictive density. At the time of writing, approximate PSIS-LOO-CV using Laplace approximation (or variational inference) and subsampling is available for any \texttt{rstan} model \citep{RStan2020}, making the method widely applicable for psychological research. Due to the flexibility of the probabilistic programming language Stan, models can be customised easily to researcher's needs \citep{Carpenter2017, Kelter2020}. 
 
 In summary, approximate LOO-CV based on posterior approximation and subsampling is an attractive method for mathematical psychologists who aim at comparing several competing statistical models under the $\mathcal{M}$-open assumption, when the number of model parameters and the sample size is large.

\section*{Declaration of interests}
The author declares no conflicts of interest.

\bibliography{library}

\end{document}